\documentclass[a4paper,11pt]{article}
\usepackage{jheppub}
 \usepackage{subfigure}

\title{Neutrino mass mechanisms from a nonstandard Higgs Lagrangian and implications for flavor hierarchies}
	\author[a]{Suppanat Supanyo$^{1}$}
 \author[b]{Chanon Hasuwannakit $^2$}
 		\author[a]{Sikarin Yoo-Kong$^1$}
   \author[c]{Lunchakorn Tannukij$^3$}
 	\affiliation[a]{The Institute for Fundamental Study (IF), Naresuan University, Phitsanulok 65000, Thailand}	
\affiliation[b]{Department of Physics, University of Oxford, Oxford OX1 3PU, United Kingdom}
  \affiliation[c]{Department of Physics, School of Science, King Mongkut's Institute of Technology Ladkrabang, Bangkok 10520, Thailand}
  \emailAdd{sikariny@nu.ac.th}
  \abstract{
      We present an alternative framework to establish the neutrino mass scale from the Higgs mechanism in a minimalist approach, which does not introduce new scalar bosons or extend the symmetry group of the standard model (SM). A nonstandard form of the Higgs Lagrangian, constructed via the inverse problem of calculus of variations, is proposed. Only one dimensionful parameter in the TeV scale is incorporated into the SM Lagrangian. The multiplicative Lagrangian model of the Higgs field plays an essential role in explaining the vast mass difference between charged fermions and Dirac neutrinos, while the Yukawa couplings for these two groups of particles naturally fall within the same scale. On the other hand, if the neutrino mass term has both Dirac and Majorana components, the mass of the mostly right-handed neutrinos in the Type-I seesaw mechanism can range from the keV scale up to slightly below the grand unification scale without requiring extremely small Yukawa couplings outside the SM regime. Furthermore, we discuss the potential of this mechanism to explain the hierarchical structure in the Yukawa couplings between first- and third-generation particles.}
  \begin{document}
 	\maketitle
  \flushbottom
\section{Introduction}\label{intro}
The origin of the neutrino mass is one of the most puzzling problems in particle physics, hinted at by the concept of naturalness, \cite{Dirac:1938mt,dijkstra2019naturalness,Williams:2015gxa,rosaler2019naturalness,Craig:2022eqo, Hook:2019mrd}, which requires no fine-tuning of parameters and no widely separated scale in the dimensionless ratios constructed from the parameters in the theory.
Based on the observations of neutrino oscillations \cite{PhysRevLett.81.1562,SNO:2002tuh,PhysRevLett.87.071301,Capozzi:2017ipn,deSalas:2020pgw} and cosmological constraints \cite{Planck:2015fie, Planck:2018vyg, KATRIN:2019yun, KATRIN:2021uub}, 
the masses of neutrinos are non-zero but appear on an unnaturally small scale compared to the masses of standard model (SM) particles.
\cite{Erler:2019hds,CDF:2022hxs,deBlas:2021wap}.  The mass ratio between the neutrino mass eigenstate ($m_{\nu_i}$) and the charged fermion mass in the SM ($m_{f_i}$) is around
\begin{align}\label{0}
   10^{-14} \lesssim\frac{m_{\nu_i}}{m_{f_i}}\lesssim 10^{-7}~.
\end{align}
   If the neutrino acquires a mass from the Higgs mechanism as in the case of the SM-charged fermions, 
 \begin{align}
   -&\mathcal{L}_\text{Yukawa}= \overline{L}_j(\lambda_{l})_{ij}\phi~ l_{iR}+\overline{Q}_j(\lambda_{u} )_{ij}\tilde{\phi} ~u_{j,R}+\overline{Q}_j(\lambda_{d})_{ij}\phi~ d_{iR}+\overline{L}_j(\lambda_{\nu} )_{l_{ij}}\tilde{\phi} ~\nu_{l_iR}+\text{h.c.}~,
 \end{align}
 where $\phi$ is Higgs doublet, $\tilde{\phi}=i\tau_2\phi$, $l_i=(l_1,l_2,l_3)=(e,\mu,\tau)$, $\nu_{l_i}=(\nu_{e},\nu_{\mu},\nu_{\tau})$, $u_j=(u_1,u_2,u_3)=(u,c,t)$, $d_i=(d_1,d_2,d_3)=(d,s,b)$, $L=(\nu~~l)^T$ and $Q=(u~~d)^T$ are left-handed lepton doublet and quark doublet, respectively, this framework gives rise to the naturalness problem.
 The Yukawa couplings ($\lambda_{i}=\sqrt{2}m_{i}/v$) of the neutrinos must be much smaller than those of the other SM fermions, 
 \begin{align}\label{013}
     10^{-14} \lesssim\frac{\lambda_{\nu_i}}{\lambda_{f_i}}\lesssim 10^{-7}~.
 \end{align}
 On the other hand, without including the neutrino into the SM, the Yukawa couplings of the fermions can be compacted within the range 
 \begin{align}\label{OSM}
     10^{-5}\lesssim\frac{\lambda_{f_i}}{\lambda_{f_j}}\lesssim 1~,
 \end{align}
 where $\lambda_{f_i}<\lambda_{f_j}$.
Obviously, the Yukawa coupling of neutrino is vastly different from those of the other SM particles. Theoretically, one can hypothesize that the neutrino does not acquire a mass from the Higgs mechanism in the same way as SM-charged fermions. 
Hence, the unnatural smallness of the neutrino mass requires an explanation from the physics beyond the SM. A prominent framework for explaining this discrepancy is the seesaw mechanism \cite{Gell-Mann:1979vob, Yanagida:1980xy, PhysRevLett.44.912}. In the Type-I version of this scheme, since the neutrino is a neutral particle, the mass term of neutrino can mix between Dirac and Majorana types 
\begin{align}\label{seesaw0}
    -\mathcal{L}^M =& m_{D} (\overline{\nu ^c}{}_L {\nu^c} _R ) + M_{R} (\overline{\nu ^c}{}_R \nu _R) + \text{h.c.} ~,
\end{align}
where the flavor index is ignored, $m_D$ denotes the Dirac mass coming from the Higgs mechanism, and $M_R$ refers to the Majorana mass arising from an extension of the SM. Here, the observed neutrino states with the diagonalization of the mass matrix can be obtained by the transformations
\begin{align}\label{mixingstate}
    &\nu_1 = i\nu_L\cos\theta_\nu  - i\nu_R^c\sin\theta_\nu~,~\nu_2 = \nu_L \sin\theta_\nu +  \nu_R^c\cos\theta_\nu~,
\end{align}
where $\theta_\nu$ is the mixing angle and $ \tan 2\theta_\nu = 2m_D/M_R $. The eigenvalues of the neutrino mass eigenstates are mixed between both types 
\begin{align} \label{SeesawSol}
    m_{1,2}=\frac{1}{2}M_R\mp \frac{1}{2}\sqrt{M_R^2+4m_D^2}~.
\end{align}
For $\nu_1$ to be mostly left-handed with a light mass and $\nu_2$ to be mostly right-handed with a heavy mass, a small mixing angle  (or $m_D \ll M_R$) is required.
Consequently, $\nu_1$ with mass $m_1$ is interpreted as an active left-handed neutrino, and $\nu_2$ with mass $m_2$ is interpreted as a heavy right-handed neutrino. 
Hence, if $M_R$ is expected to be in the TeV scale to the grand unification scale, then
$m_1$ could naturally be small in the eV scale, $m_1\simeq m_D^2/M_R\sim\mathcal{O}(1)$~eV, while $m_D$ is restricted to around a mass of the SM particles.

Even though the model could explain the smallness of the left-handed neutrino,  the right-handed mass parameter remains significantly larger than the electroweak (EW) scale, which could be seen as a lack of naturalness, implying that new physics might be needed to explain the Majorana mass term. One can assume that $M_R$ is obtained from the physics beyond the SM, for example, the SO(10) grand unification theory (GUT) \cite{Ellis:1979fg,BARBIERI1981433,Campos:2014lla,Anderson:1993fe,BABU1995365}. However, the possibility of $M_{R}<\text{TeV}$ is not completely ruled out by various experimental observations, such as a neutrino oscillation anomaly \cite{LSND:1996ubh,LSND:2001aii,PhysRevD.72.033005}, higher intensity collider \cite{Gorbunov:2007ak,Ruchayskiy:2011aa,Gninenko:2012anz,Quintero:2011yh,Helo:2011yg}. 
 In addition, the heavy right-handed neutrino with mass in this range could theoretically be the candidate for explaining the dark matter \cite{PhysRevLett.72.17,PhysRevLett.82.2832,Dolgov:2000ew, Gelmini:2004ah,Asaka:2005an,PhysRevD.107.123013} and baryogenesis \cite{Asaka:2005pn,Asaka:2006nq,Asaka:2006ek}. Therefore, if the right-handed neutrino mass is 
 experimentally concluded to be below a few GeV, these various unexplained phenomena can be solved while the Type-I seesaw mechanism could not be an appropriate resolution of the neutrino mass problem, since the Yukawa couplings of the neutrinos are still many order of magnitude smaller than the charged fermions. To keep working with the concept of naturalness in explaining unnaturally small $\lambda_\nu$ and $M_R$, more complicated structures and a new degree of freedom must be integrated into the SM, for example, the inverse seesaw model \cite{Mohapatra:1986bd,Gonzalez-Garcia:1988okv,Deppisch:2004fa}. Moreover, the type of neutrino (Dirac or Majorana) is still indistinguishable based on the current observation data from the neutrinoless double beta decay ($0\nu\beta\beta$) experiment \cite{PhysRevD.25.2951,Giuliani:2012zu,Rodejohann:2011mu,Bilenky:2012qi,Blennow:2010th,DellOro:2016tmg}. Hence, the possibility of the Dirac neutrino has not yet been ruled out. 
 In the future experimental search with a higher sensitivity of $0\nu\beta\beta$ \cite{BARABASH2015416}, if the type of neutrino is confirmed as the Dirac particle, the seesaw mechanism would no longer serve as an explanation of an unnaturally small neutrino mass. Therefore, the Dirac neutrino mass problem remains elusive, calling for an entirely new theoretical perspective.

In this paper, the objective is to reduce the widely separated scale among the SM parameters. The objective is separated into 2 major goals. The first one is to propose a mechanism to reduce the Yukawa coupling of Dirac neutrino in Eq.~\eqref{013} into the SM range
\begin{align}\label{YukawaO1}
    10^{-5}<\frac{\lambda_{\nu_i}}{\lambda_{f_j}}\lesssim 1~,
\end{align}
while a neutrino mass naturally exists in the sub-eV scale. The Dirac neutrino therefore can acquire a mass from the Higgs mechanism without the widely different scale between the Yukawa couplings of the charged fermions and neutrinos. For the second goal, if neutrinos are Majorana, including the right-handed singlet component, our framework allows the active left-handed neutrino mass $m_1$ to be in sub-eV, while the mostly right-handed neutrino mass $m_2$ spans a wider range
\begin{align}\label{MR_intro}
    1\text{keV}\lesssim m_2\lesssim 10^{12}\text{GeV}~,
\end{align}
without requiring unnaturally small(or large) $\lambda_\nu/\lambda^{\text{SM}}_f$  and $M_R$. To achieve these goals, we extend the SM in a minimalist way, which introduces neither an extension of gauge symmetry nor an addition of new scalar bosons. We will apply the nonstandard Lagrangian of the complex scalar field, the so-called multiplicative form constructed by the inverse problem of the calculus of variations \cite{douglas1941solution,hojman1981inverse,WSarlet_1982} derived in our previous work  \cite{PhysRevD.106.035020},
\begin{align}\label{Lm}
\mathcal{L}^\Lambda_\phi=F(\partial_\mu\phi^\dagger\partial^\mu\phi)f(\phi^\dagger\phi)~.
\end{align}
In this work, the Higgs Lagrangian
\begin{align}
\mathcal{L}_\text{Higgs}=(D_\mu \phi)^\dagger D^\mu \phi -V~,
\end{align}
in the SM
 \begin{align}
     \mathcal{L}_\text{SM}= \mathcal{L}_\text{Higgs}+\mathcal{L}_\text{Gauge}+ \mathcal{L}_\text{Lepton}+ \mathcal{L}_\text{Yukawa}+...~,
 \end{align}
will be replaced by the multiplicative Lagrangian
\begin{align}
    \mathcal{L}_\text{Higgs}=(D_\mu \phi)^\dagger D^\mu \phi-V \to \mathcal{L}_\text{Higgs}^\Lambda =F((D_\mu \phi)^\dagger D^\mu \phi ) f(V)~.
\end{align}
Only one dimensionful parameter, $\Lambda$, coming from the multiplicative Lagrangian will be added into the SM to explain the widely separated scale between the neutrino and the charged fermion mass. In brief, this Lagrangian is motivated by the nonuniqueness principle of Lagrangian. In classical mechanics, there are various forms of the Lagrangian instead of $T-V$ that can provide the equation of motion (EOM) as the Newton, or relativistic motion of particle \cite{degasperis2001newton,surawuttinack2016multiplicative,el2013quantum}, so the form of Lagrangian is not unique. It is plausible to hypothesize that the Lagrangian in the field theory is not restricted only to the standard $T-V$ form. Various forms of Lagrangian in the field theory could yield the EOMs that are either exactly or approximately equivalent to the Klein-Gordon and the Dirac equation \cite{supanyo2024nonstandard, PhysRevD.106.035020}.  Indeed, the framework of the nonstandard Lagrangian is not new in theoretical physics, pioneered in cosmology from both bottom-up \cite{ARMENDARIZPICON1999209,PhysRevD.63.103510,PhysRevD.103.043518} and top-down \cite{Alishahiha:2004eh} perspectives within the effective field theory.
In this paper, we stay in the bottom-up perspective. Motivated by the nonuniqueness principle, we hypothesize that the standard $T-V$ form of the complex Klein-Gordon Lagrangian is a low energy effective theory from the multiplicative Lagrangian.  The explicit expression of Lagrangian is not specifically proposed by hand but will be derived through the inverse problem of the calculus of variations.

This paper is organized as follows. In section~\ref{review}, the multiplicative form of the complex scalar field Lagrangian will be introduced. In section~\ref{smm}, we apply this Lagrangian to the SM. Furthermore, the simple constraint from the observation such as $g_{hhVV}/g_{hhVV}^\text{SM}$ is also discussed to provide the constraint of the model. In section~\ref{section4}, we apply the multiplicative Lagrangian to the neutrino mass problem and show that a mass of the Dirac neutrinos could naturally fall within the eV scale. In section~\ref{Majorana}, if the neutrinos are Majorana, this mechanism also extends the Majorana mass range in the Type-I seesaw mechanism down to eV scale without too small Yukawa coupling of neutrino and unnatural small(large) $M_R$ parameter. In section~\ref{section5}, we discuss the constraints of Dirac neutrino, the flavor problem, the validity of the model, and the open problem. Finally, in section~\ref{lastsection}, we provide the conclusion for this paper.

\section{Review on the Multiplicative Form of the complex scalar field Lagrangian}\label{review}
 We introduce an ansatz form of the complex scalar field Lagrangian, 
\begin{align*}
\mathcal{L}^\Lambda_\phi=F(\partial_\mu\phi^\dagger\partial^\mu\phi)f(\phi^\dagger\phi)~.
\end{align*}
Here, the $F$ and $f$ are unknown functions and the explicit form of $\mathcal{L}^{\Lambda}_\phi$ is derived from the combination of the two fundamental ideas in the Lagrangian mechanics:  ``the effective field perspective'' \cite{BECHTLE2022129,Brivio:2017vri,PhysRevD.104.015026,PhysRevLett.121.111801}, and ``the inverse problem of the calculus of variations'' \cite{douglas1941solution,hojman1981inverse,WSarlet_1982}.

$\bullet$ Effective field theory: Non-renormalized theory with the cutoff $\Lambda$ must be reducible into the renormalized one under the energy scale far below $\Lambda$. For example, in the non-standard Lagrangian approach such as Dirac-Born-Infeld \cite{Alishahiha:2004eh}, k-essence\cite{ARMENDARIZPICON1999209,PhysRevD.63.103510,PhysRevD.103.043518}, every modified Lagrangian of the scalar field can be reduced to the Klein-Gordon Lagrangian in the appropriate limit to describe the relativistic motion of scalar field.

$\bullet$ Inverse problem of the calculus of variations: According to the effective field framework, the form of the scalar field Lagrangian can be set into arbitrary function, and the explicit expression leading to the Klein-Gordon equation can be given by inverse calculating the Euler-Lagrangian equation. We follow the method from \cite{PhysRevD.106.035020,surawuttinack2016multiplicative}.

 By substituting \eqref{Lm} into the Euler-Lagrangian equation,
\begin{align}\label{KG0}
    0=\frac{\partial \mathcal{L}}{\partial\phi}-\partial_\mu\frac{\partial \mathcal{L}}{\partial\partial_\mu\phi}~,
\end{align}
we obtain
\begin{align}\label{el2}
   0=F\frac{\partial f}{\partial\phi}-\partial_\mu\partial^\mu\phi^\dagger f\frac{\partial F}{\partial X}-X\frac{\partial f}{\partial\phi}\frac{\partial F}{\partial X}-\partial_\mu\phi^\dagger\partial^\mu\phi^\dagger\frac{\partial f}{\partial \phi^\dagger}\frac{\partial F}{\partial X}-f\partial^\mu\phi^\dagger\partial_\mu X\frac{\partial^2 F}{\partial X^2}~.
\end{align}
 where $X=\partial_\mu\phi^\dagger\partial^\mu\phi$. Then, by assuming that the EOM \eqref{el2} can be approximated into the Klein-Gordon equation
\begin{align}\label{KGapp}
    \partial_\mu\partial^\mu\phi^\dagger+\frac{\partial V}{\partial\phi}\simeq 0~,
\end{align}
there are two terms with coefficients $\partial_\mu\phi^\dagger\partial^\mu\phi^\dagger$ and $\partial^\mu\phi^\dagger\partial_\mu X$ that cannot be reduced into the Klein-Gordon equation. The last term in Eq.~\eqref{el2} must vanish in the low momentum limit so one can set the condition $\frac{\partial^2 F}{\partial X^2}=0$. The fourth term is interpreted as the self-interaction term, which does not contribute to Eq.~\eqref{KGapp}. Then, substituting $\partial_\mu\partial^\mu\phi^\dagger$ from Eq.~\eqref{KGapp} into Eq.~\eqref{el2} and solving the inverse problem of the calculus of variations with the help of the separation variable method on the first three terms of Eq.~\eqref{el2}, the expressions for $F$ and $f$ are given by
\begin{align}
    F=&\left(\epsilon\Lambda^4+\partial_\mu \phi^\dagger \partial^\mu\phi\right)~,
    \\
    f=&e^{-\frac{V}{\epsilon\Lambda^4}}~.
\end{align}
The $\Lambda$ is a dimensionful parameter with $[\Lambda]=1$ and $\epsilon=\pm 1$ is a dimensionless parameter arising from the separation variable method.
The multiplicative form of the complex scalar field Lagrangian is then obtained in the form
\begin{align}\label{phicomplex}   \mathcal{L}^{\Lambda}_\phi=\left(\epsilon\Lambda^4+\partial_\mu \phi^\dagger \partial^\mu\phi\right)e^{-\frac{V}{\epsilon\Lambda^4}}~,
\end{align}
and the EOM itself is
\begin{align}\label{EOM}
    0=e^{-\frac{V}{\epsilon\Lambda^4}}\left(\partial_\mu\partial^\mu\phi^\dagger+\frac{\partial V}{\partial\phi}-\frac{\partial_\mu\phi^\dagger\partial^\mu\phi^\dagger}{\epsilon\Lambda^4}\frac{\partial V}{\partial \phi^\dagger}\right)~.
\end{align}
Here, we can see that, in the limit $V\ll \Lambda^4$ and $\partial_\mu\phi ^\dagger\partial^\mu\phi\ll\Lambda^4$, the Lagrangian \eqref{phicomplex} can be approximated into the Klein-Gordon Lagrangian
\begin{align}\label{phicomplex2}  
\mathcal{L}^{\Lambda}_\phi=\epsilon\Lambda^4+\partial_\mu\phi ^\dagger\partial^\mu\phi-V+\mathcal{O}(\Lambda^{-4})~.
\end{align}
The EOM \eqref{EOM} hence reduces to the standard Klein-Gordon equation \eqref{KGapp} in this limit. Therefore, this Lagrangian can be an alternative description for the relativistic motion of the complex scalar field in the low energy limit.

\section{SM with the multiplicative form of the Higgs Lagrangian}\label{smm}
In this section, as the primary focus on the neutrino problem, we concentrate on the electroweak sector including the Higgs, weak boson, and lepton sectors. The quark and gluon Lagrangians are omitted in this section.

\begin{figure}[t]
    \centering
    \subfigure[]{\includegraphics[width=0.48\linewidth]{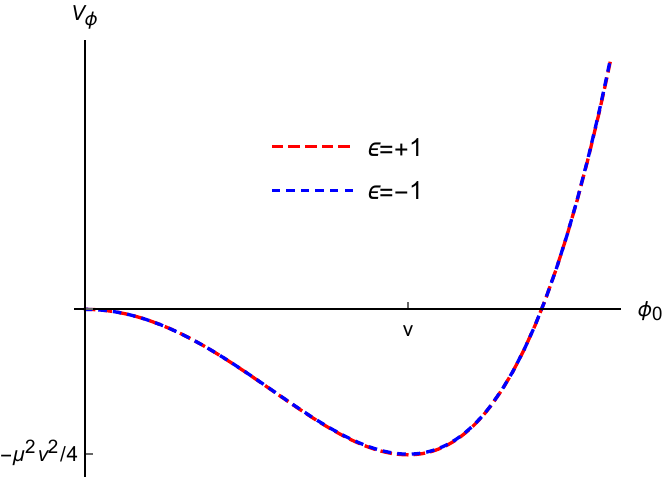}}
     \subfigure[]{\includegraphics[width=0.42\linewidth]{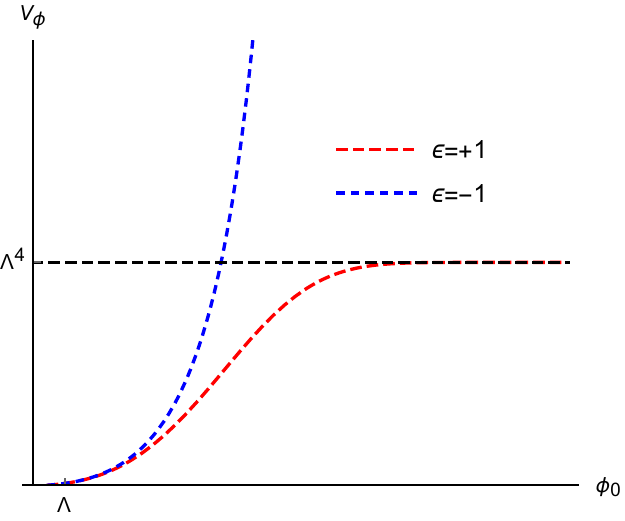}}
    \caption{(a) Higgs potential in Eq.~\eqref{Veff} in the regime $0<\phi_0\lesssim v$ and (b) Higgs potential in Eq.~\eqref{Veff} in the regime $\phi_0\gg v$.}
    \label{fig:h}
\end{figure}

To interpret the Lagrangian \eqref{phicomplex} as the Higgs field in the SM, the spacetime derivative in Eq.~\eqref{phicomplex} is promoted into the $SU(2)_L\times U(1)_Y$ covariant derivative 
\begin{align}\label{Llambda}
\mathcal{L}^{\Lambda}_\text{Higgs}=\left(\epsilon\Lambda^4+D_\mu \phi^\dagger D^\mu\phi\right)e^{-\frac{V}{\epsilon\Lambda^4}}-\epsilon \Lambda^4~,
\end{align}
where $\phi^\dagger=(0 \quad \phi_0)^\dagger/\sqrt{2}$ is a Higgs doublet in the unitary gauge which the Goldstone mode is ignored. We have subtracted the term $\epsilon \Lambda^4$ from the Lagrangian to prevent a large contribution to the cosmological constant term. 
In this model,  the potential $V$ is the Ginzburg-Landau type as in the SM
\begin{align}
    V=-\mu^2\phi^\dagger\phi+\lambda (\phi^\dagger\phi)^2~.
\end{align}
The tree-level Higgs potential in Lagrangian~\eqref{Llambda} is now modified by an exponential term introduced through the multiplicative Lagrangian as
\begin{align}\label{Veff}
V_{\phi}=-\epsilon\Lambda^4e^{-\frac{-\mu^2\phi^\dagger\phi+\lambda (\phi^\dagger\phi)^2}{\epsilon\Lambda^4}}+\epsilon\Lambda^4~.
\end{align}
The vacuum expectation value (VEV) of Higgs field can be determined by solving for the minima of the potential \eqref{Veff}, $\partial V_\phi/\partial\phi=0$, or solving the vacuum solution of the EOM \eqref{EOM} under the condition $\partial_\mu\phi=\partial_\mu\phi^\dagger=0$. Both schemes give
\begin{align}
    0=e^{\frac{2\mu^2\phi_0^2-\lambda \phi_0^4}{4\epsilon\Lambda^4}}\phi_0\left(\phi_0^2-\frac{\mu^2}{\lambda}\right)~.
\end{align}
Since the range of the exponential factor cannot be zero, the Higgs VEV ($v$) is given by
\begin{align}
    \langle \phi_0\rangle^2=v^2=\frac{\mu^2}{\lambda}~,
\end{align}
which matches the SM result and is explicitly independent of the energy scale $\Lambda$.  Depending on the choice of $\epsilon$, there are two district shapes of the Higgs potential, shown in figure \ref{fig:h}. The distinction between $V_\phi$ with $\epsilon=+1$ and $\epsilon=-1$ becomes significant at  $\phi_0=\Lambda$. For $\epsilon=-1$, the potential diverges exponentially, while, for $\epsilon=+1$, the potential converges to an asymptotic value $\Lambda^4$, shown in figure \ref{fig:h}(b). In contrast, both cases are approximately identical in the low energy regime, see figure~\ref{fig:h}(a).
We then expand the Higgs field around the VEV as
\begin{align}\label{expandphi}
    \phi(x)=\frac{1}{\sqrt{2}}\begin{pmatrix}
        0\\
        v+\chi(x)
    \end{pmatrix}~,
\end{align}
where $\chi$ is the quantum fluctuation of the Higgs field. Substituting Eq.~\eqref{expandphi} into Eq.~\eqref{Llambda}, we find that the kinetic energy for $\chi$ is not in canonical form
\begin{align}\label{KEchi}
    A(\chi)\frac{\partial_\mu \chi\partial^\mu \chi}{2}~,
\end{align}
where 
\begin{align}\label{factorA}
    A(\chi)&=e^{ \left(\frac{\mu ^2 (v+\chi )^2 \left(v^2-2 v \chi -\chi ^2\right)}{4 \Lambda ^4 v^2 \epsilon }\right)}~.
    \end{align}
    In the limit $\chi\ll \Lambda$, Eq.~\eqref{factorA}  can be serialized as
    \begin{align}
        &A(\chi)=e^{\frac{\mu ^2 v^2}{4 \epsilon\Lambda ^4  }}-\frac{\mu ^2  e^{\frac{\mu ^2 v^2}{4\epsilon \Lambda ^4  }}\chi ^2}{\epsilon\Lambda ^4  }-\frac{\mu ^2
   e^{\frac{\mu ^2 v^2}{4\epsilon \Lambda ^4  }}\chi ^3 }{\epsilon \Lambda ^4 v }+\frac{\mu ^2  e^{\frac{\mu ^2 v^2}{4  \epsilon\Lambda ^4 }}
   \left(2 \mu ^2 v^2-\epsilon \Lambda ^4 \right)\chi ^4}{4\epsilon ^2 \Lambda ^8 v^2 }
   +O\left(\chi ^5\right)~.
\end{align}
To reorganize Eq.~\eqref{KEchi} into a canonical renormalized form, one employs the field redefinition
\begin{align}\label{redchi}
    \chi=\frac{h}{\sqrt{e^{\frac{\mu ^2 v^2}{4 \epsilon
   \Lambda ^4 }}}}+\frac{\mu ^2h^2}{6\epsilon \Lambda ^4   \left(e^{\frac{\mu ^2 v^2}{4 \epsilon\Lambda ^4  }}\right)^{3/2}}+O\left(h^4\right)~.
\end{align}
The SM Lagrangian can then be reorganized as
\begin{align}    \mathcal{L}=\mathcal{L}_\text{Higgs}+\mathcal{L}_\text{Lepton}+\mathcal{L}_\text{Yukawa}+... ~,
\end{align}
where
\begin{align}
    \mathcal{L}_\text{Higgs}=&\frac{1}{2}\partial_\mu h\partial^\mu h-\frac{M_h^2}{2}h^2-\frac{\lambda_3}{3!}h^3-\frac{\lambda_4}{4!}h^4+O(h^5)~,\label{Lhiggs}
  \\
     \mathcal{L}_\text{Lepton}=&\overline{l}_{iL}\gamma^\mu i\partial_\mu l_{iL}+\overline{l}_{iR} i\gamma^\mu\partial_\mu l_{iR}+\overline{\nu}_{l_iL}\gamma^\mu i\partial_\mu \nu_{l_iL}-m_{l_i}(\overline{l}_{iL}l_{iR}+\overline{l}_{iR}l_{iL})~,\label{Lclepton}
    \\
    \mathcal{L}_\text{Yukawa}=&-y_{l_i}h (\overline{l}_{iL}l_{iR}+\overline{l}_{iR}l_{iL})+\mathcal{O}(h^2)~,\label{Lyu}
\\
    \mathcal{L}_\text{Gauge}=&-\frac{1}{4}F_{\mu\nu}F^{\mu\nu}-\frac{1}{4}Z_{\mu\nu}Z^{\mu\nu}+\frac{M_Z^2}{2}Z_\mu Z^\mu-\frac{1}{2}W^+_{\mu\nu}W_-^{\mu\nu}+M_W^2W^{\mu}_+W_{\mu}^-,
    \\
    \mathcal{L}_{llV}=&-g_{ll\gamma} A_\mu (\bar{l}_{iL} \gamma^\mu l_{iL}
    +\bar{l}_{iR}\gamma^\mu l_{iR})+g_{l\nu W}( W^+_\mu \overline{\nu}_{l_iL}\gamma^\mu l_{iL}+ W^-_\mu \overline{l}_{iL}\gamma^\mu \nu_{l_{i}L})
    \nonumber
    \\
   & +g_{\nu\nu Z} Z_\mu\overline{\nu}_{l_iL}\gamma^\mu\nu_{l_iL}-g_{llZ,L} Z_\mu \overline{l}_{iL}\gamma^\mu l_{iL}+g_{llZ,R}\overline{l}_{iR}\gamma^\mu l_{iR},\label{llV}
    \\
    \mathcal{L}_{hVV}=& g_{hWW}h W^\mu_+W_\mu^-+\frac{g_{hZZ}}{2}hZ^2~,
    \\
    \mathcal{L}_{hhVV}=& \frac{g_{hhWW}}{2}h^2 W^\mu_+W_\mu^-+\frac{g_{hhZZ}}{4}h^2Z^2~.
\end{align}
The gauge boson and charged lepton masses can be expressed in terms of the Higgs VEV and the coupling constant as follows,
\begin{align}
    &M_W^2=\frac{1}{4} g^2 v^2 e^{\frac{\mu ^2 v^2}{4 \epsilon\Lambda ^4}}~,\label{Wmass}
    \\
    &M_Z^2=\frac{1}{4} g^2 v^2 \sec ^2(\text{$\theta $w}) e^{\frac{\mu ^2 v^2}{4\epsilon \Lambda ^4}}~,\label{Zmass}
    \\
    &m_{l_i}=\frac{\lambda_{l_i}v}{\sqrt{2}}~,\label{massfSM}
    \\
    &M_h^2=2 \mu ^2~, \label{Hmass}
\end{align}
while the photon mass vanishes, $m_\gamma=0$, and $M_h=125~\text{GeV}$\cite{HiggsMass125,HiggsMass1252}. Here, we note that the quadratic Lagrangians of quark in the mass eigenstate are similar to those of charged leptons, and the quark mass is also written in the same form as Eq.~\eqref{massfSM}.  From Eq.~\eqref{Wmass}-\eqref{Zmass}, the tree level custodial symmetry, defined from the parameter $\rho_\text{tree}=M_W^2/M_Z^2\cos^2\theta_W$, obviously ensures the SM result $\rho_\text{SM,tree}=1$ \cite{VELTMAN197789,Sikivie:1980hm,COSTA1988244}.  In addition, the coupling constants of the three-legged interaction terms are
\begin{align}\label{3leggedcoupling}
   & \lambda_3=\frac{3 M_h^2}{v}e^{-\frac{ M_h^2v^2}{16 \epsilon  \Lambda ^4}}~,~ y_{l_i}=\frac{\lambda _{l_i}}{\sqrt{2}}e^{\frac{M_h ^2 v^2}{16 \epsilon\Lambda ^4 }}~,~
   g_{ll\gamma}=e~,~g_{l\nu W}=\frac{g}{\sqrt{2}}~,~g_{llZ,L}=e\cot2\theta_W~,\nonumber
   \\
   &g_{llZ,R}=e\tan\theta_W~,~g_{\nu\nu,Z}=e\csc 2\theta_W,~g_{hWW}=\frac{g^2 v}{2} e^{\frac{M_h^2 v^2 }{16 \epsilon\Lambda ^4  }} ~,~g_{hZZ}=\frac{ g^2 v}{2} \sec ^2\theta_W e^{\frac{ M_h^2v^2}{16 \epsilon \Lambda ^4 }}~,
\end{align}
where $e=g\sin\theta_W$. Then, from Eqs.~\eqref{Wmass}-\eqref{Zmass}, the VEV can be written in terms of the Fermi coupling constant ($G_F=1.166\times 10^{-5}$GeV$^{-2}$  \cite{Erler:2019hds}) as
\begin{align}\label{HiggsVEV}
    v_\text{SM}^2=v^2 e^{\frac{M_h^2  v^2}{8 \Lambda ^4 \epsilon}}~,
\end{align}
or
\begin{align}\label{HiggsVEV2}
   v^2= v_{\text{SM}}^2 e^{-W\left(\frac{M_h^2 v_{\text{SM}}^2}{8 \Lambda ^4 \epsilon }\right)}~,
\end{align}
 where $W$ is the product logarithm function  also known as the Lambert $W$ function and $v_\text{SM}=1/\sqrt{\sqrt{2}G_F}=246$ GeV. We find that the value of Higgs VEV, $v$, is unidentified since the parameter $\Lambda$ is not yet determined. The VEV could vary depending on the value of the free parameter $\Lambda$ and the choice of $\epsilon$. In the case $\epsilon=+1$, we have $v<v_\text{SM}$ and, in the case $\epsilon=-1$, we have   $v>v_\text{SM}$. In both cases, in the limit $\Lambda\to \infty$, the value of $v$ asymptotically converges to $v_\text{SM}$.
By substituting  $v$ from Eq.~\eqref{HiggsVEV2} into Eq.~\eqref{3leggedcoupling}, the coupling constants of the three-legged interactions in Eq.~\eqref{3leggedcoupling} can be reorganized in terms of the SM parameters 
\begin{align}\label{3coupling}
  & \lambda_3=\frac{3 M_h^2}{v_\text{SM}},~ y_{l_i}=\frac{m_{l_i}}{v_\text{SM}},~
   g_{ll\gamma}=e,~g_{l\nu W}=\frac{g}{\sqrt{2}},~g_{llZ,L}=e\cot2\theta_W,\nonumber
   \\
   &g_{llZ,R}=e\tan\theta_W,~g_{\nu\nu,Z}=e\csc 2\theta_W,~g_{hWW}=\frac{g^2 v_\text{SM}}{2}   ,~g_{hZZ}=\frac{ g^2 v_\text{SM}}{2} \sec ^2\theta_W,
\end{align}
where we have substituted $\lambda_{l_i}=\sqrt{2}m_{l_i}/v$.
These results suggest that the tree-level predictions are insensitive to $\Lambda$. According to these coupling constants, the predicted decay rates of the Higgs into vector bosons and fermions in this model still satisfy the experimental observation from the particle collider \cite{sirunyan2019measurement,sirunyan2020measurement,aad2015evidence,sirunyan2018observation,aaboud2019cross,sirunyan2018evidence,CMS:2020xwi,CMS:2017zyp,CMS:2022urr}.

\begin{figure}[t]
    \centering
    \includegraphics[width=0.5\linewidth]{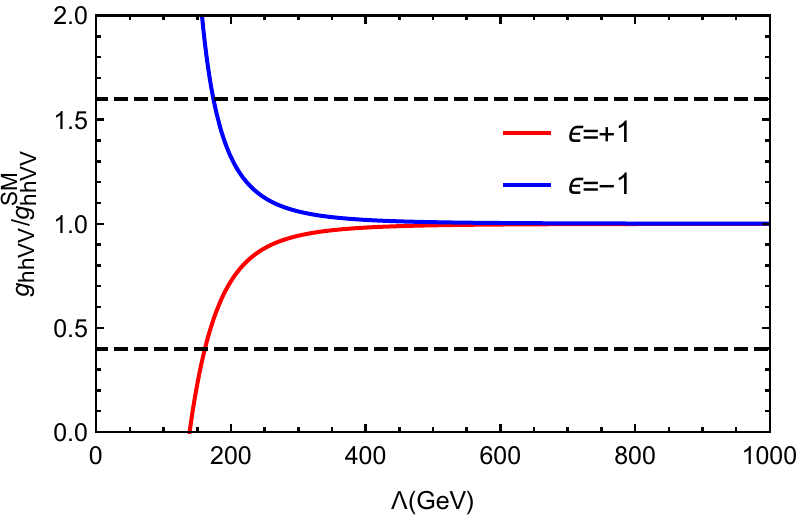}
    \caption{The ratio of $g_{hhVV}/g_{hhVV}^{\text{SM}}$ in tree-level varying with respect to the parameter $\Lambda$.}
    \label{fig:ghhvv}
\end{figure}

The deviations from the SM appear in the quartic interaction terms,
\begin{align}\label{4legged}
 &\lambda_4=\frac{3M_h^2 }{v_{\text{SM}}^2}\left(1-\frac{8}{3} W\left(\frac{M_h^2 v_{\text{SM}}^2}{8 \epsilon\Lambda ^4  }\right)\right)~,~g_{hhVV}=\frac{g^2 c_{VV}}{2} \left(1-4 W\left(\frac{M_h^2 v_{\text{SM}}^2}{8 \epsilon\Lambda ^4  }\right)\right)~,
\end{align}
while the SM predictions are
\begin{align}\label{4leggedSM}
     &\lambda_4^{\text{SM}}=\frac{3M_h^2 }{v_{\text{SM}}^2}~,~g_{hhVV}^{\text{SM}}=\frac{g^2 c_{VV}}{2}~,
\end{align}
where $VV$ can be $WW$ and $ZZ$, and $c_{WW}$=1,  $c_{ZZ}=\sec^2\theta_W$, respectively.
From the theoretical constraint of the partial wave unitary $J=0$, $\lambda_4/\lambda_{4,\text{SM}}\lesssim 65$ \cite{DiLuzio:2017tfn} is required to preserve unitarity of scattering amplitude. However,  accurately determining  $\lambda_4$ has not yet come to the conclusion  even at 100 TeV collider with high luminosity \cite{Dolan:2013rja,Dolan:2015zja,Bishara:2016kjn,PhysRevD.101.076012,Arganda:2018ftn}.  In contrast to $\lambda_4$, a recent observation from the ATLAS detector with the $95$ percent confident level \cite{ATLAS:2024ish} imposes a strong constraint on $g_{hhVV}/g_{hhVV}^{\text{SM}}$ as follows
\begin{align}
    0.4<\frac{g_{hhVV}}{g_{hhVV}^\text{SM}}<1.6~.
\end{align}
This constraint sets a lower limit on $\Lambda$  
\begin{align}\label{lowerlambda}
    \Lambda\gtrsim 170~\text{GeV},
\end{align}
 shown in figure~\ref{fig:ghhvv}. 
 On the other hand, in the limit $\Lambda^4\gg M_h^2v^2_\text{SM}/8\sim (104)^4$~GeV$^4$, the quartic couplings $\lambda_4$ and $g_{hhVV}$ in Eq.~\eqref{4legged} receive a small modification 
\begin{align}
    &\lambda_4\simeq \lambda_{4}^\text{SM}\left(1-\frac{M_h^2 v^2_\text{SM}}{3\epsilon\Lambda^4}\right)~,~g_{hhVV}\simeq g_{hhVV}^\text{SM}\left(1-\frac{M_h^2 v^2_\text{SM}}{2\epsilon\Lambda^4}\right)~.
\end{align}
Therefore, the upper limit of $\Lambda$ is unconstrained by this experimental observation.

To specify the upper value of $\Lambda$, we apply a theoretical concept of the Dirac naturalness in which the dimensionless ratios of parameters in the theory are naturally expected to be of order unity
\begin{align}\label{DiracCondition}
    \text{dimensionless ratio}\simeq \mathcal{O}(1)~,
\end{align}
where $\mathcal{O}(1)$ falls approximately within the range   $10^{-3}-1$, \cite{Dirac:1938mt,dijkstra2019naturalness}.  In this model, three dimensionless ratios are constructed from the three dimensionful parameters in our theory
\begin{align}
    \frac{\mu}{\Lambda},~\frac{v}{\Lambda},~\frac{\mu}{v}~.
\end{align}
Applying Dirac naturalness, we obtain a theoretical constraint  $88~\text{GeV}\lesssim \Lambda\lesssim 88~\text{TeV}$. Combining this theoretical constraint with the observational constraints in Eq.~\eqref{lowerlambda}, we therefore obtain the $\Lambda$ in the regime
\begin{align}\label{Lambdaconstraint}
    170~\text{GeV}\lesssim\Lambda\lesssim 88~\text{TeV}.
\end{align}
Then, we apply the multiplicative Higgs model and the constraint Eq.~\eqref{Lambdaconstraint} to study the neutrino mass problem for both Dirac and Majorana types.

\section{Mass of Dirac neutrino}\label{section4}
In this section, we demonstrate how Dirac neutrinos can acquire an unnaturally small mass compared to the charged fermions, while the Yukawa coupling remains the same order of magnitude.

\subsection{Motivation}
Modifying the Higgs Lagrangian to a multiplicative form provides flexibility in various ways to construct the gauge invariant theory. The Lagrangian is not limited to a linear combination of gauge invariant operators, instead, these operators can be incorporated both inside and outside the multiplicative structure of the Higgs Lagrangian.

The idea originates from studying the addition of an arbitrary gauge invariant operator $\hat{A}$ inside and outside the multiplicative Lagrangian \eqref{Llambda}, see the discussion in \cite{PhysRevD.106.035020}, as follows,
\begin{align}\label{mechanism}
    -\epsilon\hat{A}+\left(\epsilon\Lambda^4+D_\mu \phi^\dagger D^\mu\phi+\epsilon\hat{A}\right)e^{-\frac{V}{\epsilon\Lambda^4}}~.
\end{align}
After the EW transition, the $\hat{A}$ term in the Lagrangian \eqref{mechanism}  can be expressed as 
\begin{align}\label{A}
  \epsilon  \left(e^{W\left(\frac{M_h^2 v_{\text{SM}}^2}{8\epsilon  \Lambda ^4 }\right)}-1\right)\hat{A}~.
\end{align}
In the limit $\Lambda^4\gg M_h^2 v_\text{SM}^2$/8, the $\hat{A}$-term is suppressed by the large energy scale $\Lambda^4$ as
\begin{align}\label{A2}
  \epsilon  \left(e^{W\left(\frac{M_h^2 v_{\text{SM}}^2}{8 \Lambda ^4 \epsilon }\right)}-1\right)\hat{A}\simeq \frac{M_h^2 v_\text{SM}^2}{8\Lambda^4}\hat{A}~.
\end{align}
Under the constraint $\Lambda$ in Eq.~\eqref{Lambdaconstraint}, the coefficient of the operator $\hat{A}$ can naturally acquire a small value within the range
\begin{align}
    10^{-12}\lesssim \frac{M_h^2 v_\text{SM}^2}{8\Lambda^4}\lesssim 10^{-1}~.
\end{align}
Then, we apply this mechanism to explain the Dirac neutrino mass in the next section.

 \subsection{Dirac neutrino} \label{DiracNeutrino}
In this section, to avoid the redundancy of non-diagonal neutrino mass matrix in the interaction basis $\nu_{l_i}$, we consider the neutrino Yukawa couplings in the mass eigenstate $\nu_i$ in which the mass matrix is diagonal.

\begin{figure}[t]
    \centering
    \includegraphics[width=0.5\linewidth]{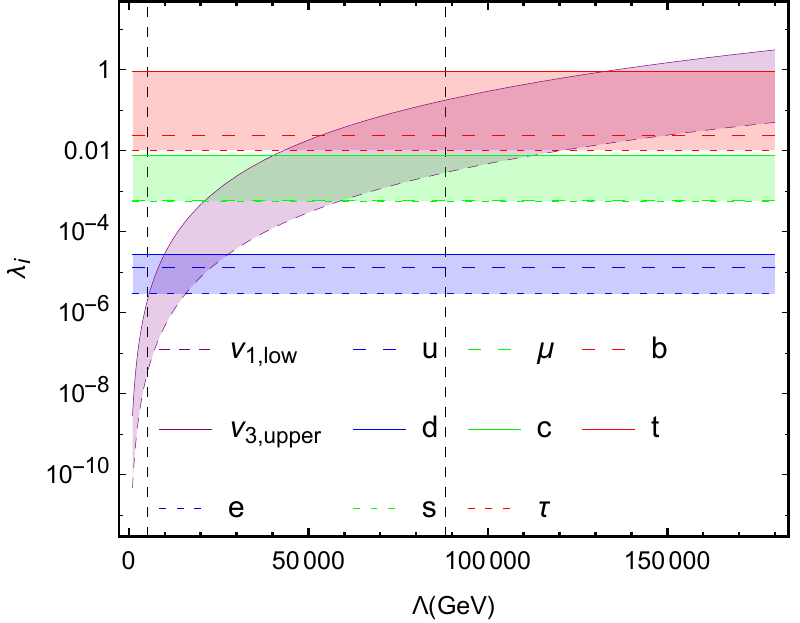}
    \caption{The plot of Yukawa couplings of the charged fermion, and neutrino from Eq.~\eqref{DiracMass} versus the scale $\Lambda$.}
    \label{fig:nuLambda}
\end{figure}

 We take an ansatz form of the Lagrangian
\begin{align}\label{Lynu}
    & \mathcal{L} \supset  + \epsilon\overline{L}_j(\lambda_{\nu} )_{l_{ij}}\tilde{\phi} ~\nu_{l_iR}+\text{h.c}.\nonumber
      \\
      &-\left(\epsilon\Lambda^4+D_\mu \phi^*D^\mu\phi-\epsilon\overline{L}_j(\lambda_{\nu} )_{l_{ij}}\tilde{\phi} ~\nu_{l_iR}+\text{h.c.}\right)e^{-\frac{V}{\epsilon\Lambda^4}}~.
\end{align}
 Here, \textit{``the Yukawa coupling terms of charged fermions are written outside the multiplicative Lagrangian.
  On the other hand, the Yukawa interactions of Dirac neutrinos with three flavors $\nu_{l_i}$ (for $l_1=e,~l_2=\mu,~l_3=\tau$) are written inside and outside the multiplicative Lagrangian."} 
Then, after the EW phase transition, the Yukawa interaction in Eq.~\eqref{Lynu} can be rewritten as
\begin{align}\label{46}
    &\epsilon \left(1-e^{-\frac{V}{\epsilon\Lambda^4}}\right)\overline{L}_j(\lambda_{\nu} )_{l_{ij}}\tilde{\phi} ~\nu_{l_iR}\nonumber
    \\
    &=\left(\frac{M_h^2 v_\text{SM}^2}{8\Lambda^4}\right)\left(\frac{(\lambda_\nu)_{l_{ij}} v_\text{SM}}{\sqrt{2}}\overline{\nu}_{l_j,L}\nu_{l_j,R}+\frac{(\lambda_\nu)_{l_{ij}}}{\sqrt{2}}h\overline{\nu}_{l_j,L}\nu_{l_i,R}+\mathcal{O}(h^2)\right)+\mathcal{O}(\Lambda^{-8})~.
\end{align}
The neutrino field in the interaction basis $\nu_{l_i}$ can be reorganized into the neutrino mass eigenstate $\nu_j$ by the unitary transformation
\begin{align}\label{massbasis}
    \nu_l= U_\nu\nu_i~.
\end{align}
Here, if we assume that the charged leptons do not oscillate,  $U_\nu=U_\text{PMNS}$ is Pontecorvo Maki Nakagawa Sakata (PMNS) matrix \cite{Maki:1962mu,Pontecorvo:1957cp,Gonzalez-Garcia:2007dlo}.
To diagonalize neutrino mass term by using Eq.~\eqref{massbasis}, the mass terms and Yukawa interactions of the neutrinos are given by
\begin{align}
    -&\mathcal{L}_{mass}=  m_{\nu_i}\overline{\nu}_{iL}\nu_{iR}+\text{h.c.}~,
    \\
    -&\mathcal{L}_{Y}= y_{\nu_i}h\overline{\nu}_{iL}\nu_{iR}+\text{h.c.}~,\label{yukawa2}
\end{align}
where 
\begin{align}\label{DiracMass}
 \quad m_{\nu_i}=\left(\frac{M_h^2 v_\text{SM}^2}{8\Lambda^4}\right)\frac{\lambda_{\nu_i}v_\text{SM}}{\sqrt{2}}~,
\end{align}
and 
\begin{align}\label{ytree}
   y_{\nu_i}=\frac{m_{\nu_i}}{v_\text{SM}}~.
\end{align}
 Now, let us consider the ratio between neutrino mass $m_{\nu_i}$ in Eq.~\eqref{DiracMass} and fermion mass $m_f$ in Eq.~\eqref{massfSM}
\begin{align}\label{Diracmassratio0}
    \frac{m_{\nu_i}}{m_{f_j}}=&\frac{M_h^2 v_\text{SM}^2}{8\Lambda^4}\left(\frac{\lambda_{\nu_i} }{\lambda _{f_j}}\right)
    ~.
\end{align}
We see that the mass of the neutrino in the basis of the mass eigenstate can be smaller than the mass of the charged fermions in the SM by many orders of magnitude, while the ratio between Yukawa couplings is of Dirac order unity $\mathcal{O}(1)$. 
To verify this argument, the predicted mass of neutrino must be consistent with the current observations. 
According to the Direct neutrino-mass measurement of the neutrino mass sum $\sum_{i} m_{\nu_i}=m_{\nu_1}+m_{\nu_2}+m_{\nu_3}$ and the neutrino oscillation constraint of  the mass square difference, $\Delta m_{ij}^2=m_{\nu_i}^2-m_{\nu_j}^2$, in the normal hierarchy (NH) mass $m_{\nu_1}< m_{\nu_2}<m_{\nu_3}$, from the particle data group \cite{ParticleDataGroup:2024cfk}
\begin{align}\label{constraint1}
 \sum m_{\nu_i}&\simeq 0.060~\text{eV}-0.12~\text{eV}~,\nonumber
 \\
     \Delta m_{21}^2&\simeq 7.53\times 10^{-5}~\text{eV}^2~,\nonumber
    \\
    \Delta m_{32}^2&\simeq 2.455\times 10^{-3}~\text{eV}^2~,
\end{align}
the neutrino masses are in the range of a few meV,
\begin{align}
  0.00096~\text{eV}\lesssim  &m_{\nu_1}\lesssim 0.030~\text{eV}~,\label{nur1}
  \\
  0.0087~\text{eV}\lesssim  &m_{\nu_2}\lesssim 0.031~\text{eV}~,\label{nur2}
  \\
  0.050~\text{eV}\lesssim  &m_{\nu_3}\lesssim 0.059~\text{eV}~,\label{nur3}
\end{align}
while, in the inverse hierarchy (IH) analyzed in appendix \ref{IHsec}, the mass ranges of neutrinos are approximately in the same order with Eqs.~\eqref{nur1}-\eqref{nur3}. Since the size of the Yukawa coupling does not deviate, we focus solely on the NH case.
By combining the constraints from Eqs.~\eqref{nur1}-\eqref{nur3} with the neutrino mass formula in Eq.~\eqref{DiracMass}, the Yukawa coupling for the neutrino mass eigenstate falls within the range
\begin{align}\label{nuYrange1}
  \frac{4.7\times 10^{-23}\Lambda^4}{\text{GeV}^4}  \lesssim\lambda_{\nu_i}\lesssim \frac{2.9\times 10^{-21}\Lambda^4}{\text{GeV}^4},
\end{align}
where the lower and upper limits correspond to $\lambda_{\nu_1}$ and $\lambda_{\nu_3}$, respectively. The plot of the Yukawa coupling $\lambda_\nu$ varying on the scale $\Lambda$ can be illustrated in figure~\ref{fig:nuLambda}.
From Eq.~\eqref{nuYrange1}, for $\Lambda\gtrsim 373$ TeV, we have $\lambda_{\nu_1}>\lambda_t$ so the Yukawa couplings of neutrinos are entirely outside the SM regime. This regime also leads to a widely separated scale of the dimensionful parameter, $\Lambda\gg \mu$ and $\Lambda\gg v$, without a theoretical explanation. Thus, it is technically unnatural. For $\Lambda>5.7~\text{TeV}$, $\lambda_\nu$ can fall within the SM range, see the intersection between the purple and blue bands in figure~\ref{fig:nuLambda}. 
Under the naturalness constraint $\Lambda<88$ TeV, $\lambda_\nu$ can be located on the SM scale. However, it cannot reach the value of $\lambda_t$. At the $\Lambda=88$ TeV, $\lambda_{\nu}\sim 7\lambda_b$. Consequently, the Yukawa couplings of neutrinos can range between  
\begin{align}
    \lambda_e\lesssim\lambda_\nu\lesssim  7\lambda_b~,
\end{align}
depending on the scale $\Lambda$. This results in the Yukawa coupling ratios of the neutrino to the other SM fermion  naturally  falling within the SM regime 
\begin{align}\label{ratiolambda}  10^{-5}\lesssim\lambda_{\nu_i}/\lambda_{f_j}\lesssim 1~,
\end{align}
therefore, there is no dissimilarity in the Yukawa couplings of neutrinos and the rest of the SM fermions. Consequently, the sub-eV mass scale of the three generations of the active neutrinos can be naturally obtained if the neutrino Yukawa interaction is written inside and outside the multiplicative Lagrangian \eqref{Lynu}.

\section{Modified Type-I seesaw mechanism}\label{Majorana}

In this section, we consider the contribution from the right-handed Majorana mass term, which does not violate the electroweak symmetry. We will show that the existence of the Higgs multiplicative Lagrangian does not conflict with the Type-I seesaw mechanism. Moreover, the multiplicative Lagrangian can potentially extend the range of the right-handed Majorana neutrino mass to below the TeV scale while keeping the ratio of the Yukawa coupling between neutrino and charged fermion around of order unity $\mathcal{O}(1)$.

From the previous section, we have proposed two possible ways to introduce the Yukawa interaction terms into the Lagrangian. One is to include the terms outside and the other is to add the terms both inside and outside of multiplicative Lagrangian. 
When the seesaw mechanism is implemented, our Lagrangian is required to contain both Dirac and Majorana mass terms. The Dirac mass term comes from the Higgs mechanism while the Majorana mass could come from the extended part of the SM, for example, the SO(10) grand unification theory \cite{Ellis:1979fg,BARBIERI1981433,Campos:2014lla,Anderson:1993fe,BABU1995365}. The Lagrangian can be expressed as
\begin{align}\label{Lynu2}
    & \mathcal{L}\supset +\epsilon\overline{L}_j(\lambda_{D} )_{l_{i,j}}\tilde{\phi} ~\nu_{l_iR}-M_{l_i,R}\overline{\nu^c_{l_i,R}}\nu_{l_i,R}+\text{h.c.}\nonumber
      \\
      &+\left(\epsilon\Lambda^4+D_\mu \phi^*D^\mu\phi-a\epsilon \overline{L}_j(\lambda_{D} )_{l_{i,j}}\tilde{\phi} ~\nu_{l_iR}+\text{h.c.}\right)e^{-\frac{V}{\epsilon\Lambda^4}}~,
\end{align}
 where $a=0$ or $1$. Here, we replace $\lambda_\nu$ with the notation $\lambda_D$ to match with the standard notation of the Type-I seesaw mechanism.
We will consider two distinct schemes for introducing $m_D$ either only outside ($a=0$) or inside-outside ($a=1$) the multiplicative Lagrangian, leading to two different results for $m_D$ presented in table~\ref{tab:mDMR0}.
\begin{table}[h!]
    \centering
    \begin{tabular}{c|c|c}
        & Yukawa interaction  &  $m_D$ 
        \\
        & (Dirac Mass term) & 
         \\
         \hline
     1&  out ($a=0$) &  $\frac{\lambda_{D} v}{\sqrt{2}}$ 
       \\
       \hline
     2&  in-out ($a=1$) & $\left(\frac{M_h^2 v_\text{SM}^2}{8\Lambda^4}\right)\frac{\lambda_{D} v}{\sqrt{2}}$ 
    \end{tabular}
\caption{This table shows the expression of $m_D$  in 2 different schemes, where the abbreviations ``out'' and ``in-out'' mean that the Yukawa interaction is written ``outside'' of the multiplicative Lagrangian, and written both ``inside and outside'' out the multiplicative Lagrangian, respectively.}
    \label{tab:mDMR0}
\end{table}
We note that, in order to simplify the mass scale of neutrinos from the seesaw mechanism, we consider three flavor neutrinos with the diagonal $3\times 3$ Dirac and Majorana mass matrix. As a consequence, each generation of neutrinos can be diagonalized separately with left-handed and right-handed mass eigenstates.
The mass matrix then can be diagonalized giving two mass eigenstates as in Eq.~\eqref{SeesawSol}. 
Then, the effect of each scheme on active neutrino and heavy right-handed neutrino mass is analyzed below.

$\bullet$ In the first scheme ($a=0$), one can see that the mass $m_D$ is naturally in the same order of magnitude as $m_f$. However, $M_R$ is an arbitrary value coming from the physics beyond SM. 
Obviously, there is no effect from the multiplicative Lagrangian on the mass of the neutrinos and the result follows the standard Type-I seesaw mechanism.
If the small mixing angle is assumed ($m_D \ll M_R$), the mass eigenstates split into active neutrino $m_1$ and heavy right-handed neutrino $m_2$ with
\begin{align}\label{seesawsolsmallmixing}
    m_1\simeq \frac{m_D^2}{M_R},~m_2\simeq M_R,
\end{align}
respectively. 
Here, the mass $m_1$ can be lifted to a small value as $M_R$ becomes very large in the Type-I seesaw mechanism.
If $m_D \simeq m_e$, $M_R$ is constrained to be around 2 TeV providing active neutrino mass in the sub-eV scale. On the other hand, for $m_D$ in the EW scale, $M_R$ is required to be around the GUT scale. In order to achieve active neutrino with sub-eV mass, right-handed Majorana mass needs to be very large,
\begin{align}
    10^3~\text{GeV}\lesssim M_R\lesssim 10^{15}~\text{GeV}.
\end{align}
Although the evidence for a heavy right-handed neutrino has not yet been found, the possibility of a heavy right-handed neutrino with a mass around or below GeV is not ruled out. 
This scheme can still be an explanation but the Yukawa coupling constants fall outside the SM regime $\lambda_{D}/\lambda_f \ll 10^{-6}$.
\begin{figure}[t]
    \centering
    \subfigure[]{ \includegraphics[width=0.45\linewidth]{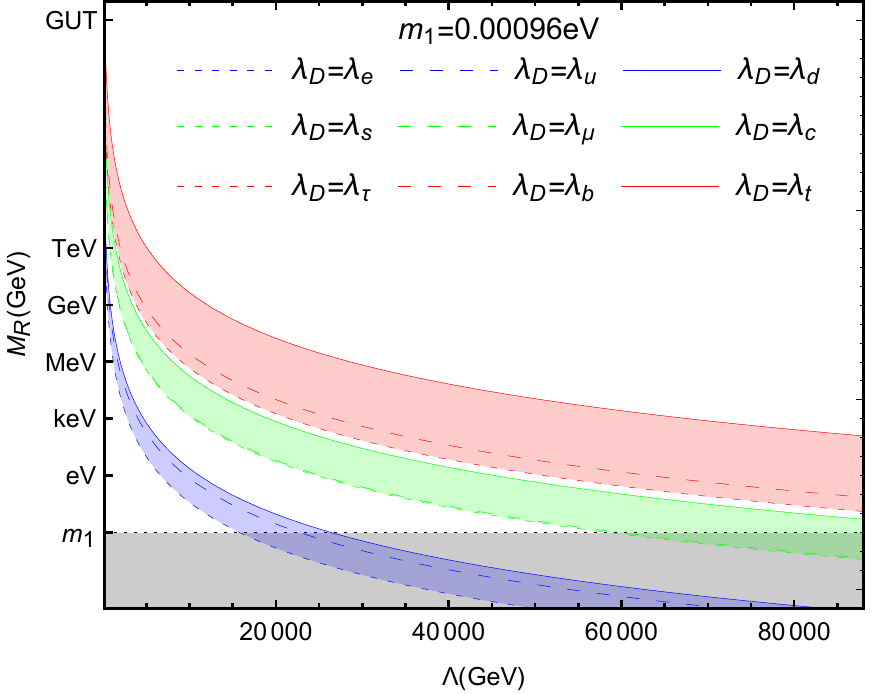}}
   \subfigure[]{ \includegraphics[width=0.45\linewidth]{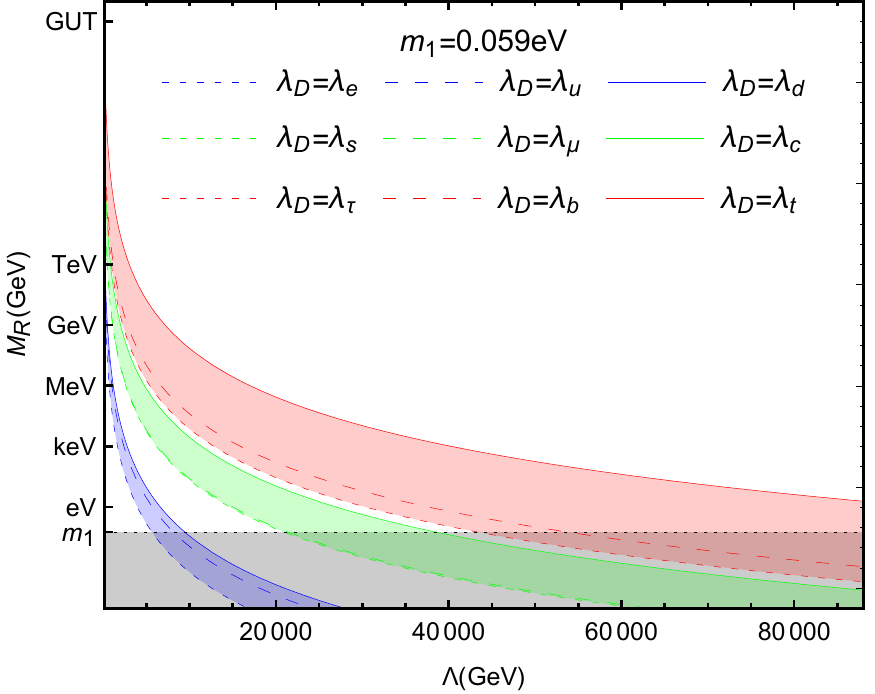}}
    \caption{The plot of $M_R$ from Eq.~\eqref{Mrx} varying with the scale $\Lambda$, where (a) $m_1=\text{min} (m_{\nu_i})$, (b) $m_1=\text{max} (m_{\nu_i})$. The gray shading area refers to the large mixing regime.}
    \label{fig:Mr}
\end{figure}

$\bullet$ In the second scheme ($a=1$), since the Dirac mass term $m_D$ is suppressed by a factor of $M_h^2 v_\text{SM}^2 / 4\Lambda^4$, 
the mixing angle is heavily suppressed by the parameter $\Lambda$ 
\begin{align}
    \tan 2\theta_\nu = \frac{M_h^2 v_\text{SM}^2}{4\Lambda^4}\frac{\lambda_{D} v_\text{SM}}{\sqrt{2} M_R} ~.
\end{align}
Then, the small mixing angle condition can be realized even for a small value of $M_R$ below the TeV scale and $M_R<\lambda_D v_\text{SM}$.
Therefore, the active neutrino and heavy neutral lepton masses are given by
\begin{align}\label{Mrx}
     m_1\simeq\left(\frac{M_h^4 v_\text{SM}^4}{128\Lambda^8}\right)\frac{\lambda_D^2v_\text{SM}^2}{M_R}, ~m_2\simeq M_R.
\end{align}
The small mixing condition is still valid, where $\tan 2\theta_\nu=2\sqrt{m_1/M_R}\ll 1$ so $ M_R\gtrsim m_1$. Here, if $\lambda_D / \lambda_f$,   $\Lambda$, and $m_1$ range in Eq.~\eqref{OSM}, Eq.~\eqref{Lambdaconstraint}, and Eqs.~\eqref{nur1}-\eqref{nur3}, respectively, we can obtain the possible mass range of the right-handed Majorana mass shown in figure~\ref{fig:Mr}. Thus, from figure~\ref{fig:Mr}(a) and figure~\ref{fig:Mr}(b), the value of $m_2$ falls within the regime
\begin{align}\label{MRr}
   1~\text{meV} \lesssim m_2\lesssim 10^{15}~\text{GeV}~,
\end{align}
without the large mixing between left and right-handed neutrino states, where $\Lambda>170$GeV.
Interestingly, the mostly right-handed neutrino mass can naturally be smaller than the EW scale to lift the active neutrino mass up to the sub-eV scale.

However, to maintain naturalness, the dimensionful parameter \( M_R \) should be within the same range as the other dimensionful parameters in the model, including $\Lambda,~\mu$, and $v$. Hence, the natural value of $m_2$ would be expected around the EW scale. To obtain $1\text{meV}\lesssim m_2\ll 100$~GeV, $M_R$ is unnaturally small. Conversely, to obtain $100~\text{GeV}\ll m_2\lesssim 10^{15}$~GeV, $M_R$ is unnaturally large.

\subsection{Natural small right-handed Majorana mass in the standard Type-I seesaw mechanism}\label{SmallMajoranamassKe}
 We can address the problem of an unnaturally small $M_R$ by incorporating both the Yukawa interaction and the right-handed Majorana mass term inside and outside the multiplicative Lagrangian as
\begin{align}\label{Lynu3}
    & \mathcal{L}\supset +\epsilon\overline{L}_j(\lambda_{\nu} )_{l_{i,j}}\tilde{\phi} ~\nu_{l_iR}+\epsilon M_R\overline{\nu^c_R}\nu_R+\text{h.c.}\nonumber
      \\
      &+(\epsilon\Lambda^4+D_\mu \phi^*D^\mu\phi-\epsilon a \overline{L}_j(\lambda_{\nu} )_{l_{i,j}}\tilde{\phi} ~\nu_{l_iR}
     -\epsilon M_R\overline{\nu^c_R}\nu_R+\text{h.c.})e^{-\frac{V}{\epsilon\Lambda^4}}~.
\end{align} 
After EW phase transition, the neutrino mass term is expressed as
\begin{align}
    \frac{M_h^2 v_\text{SM}^2}{8\Lambda^4}\left(\frac{\lambda_D v_\text{SM}}{\sqrt{2}}\overline{\nu}_L\nu_R+M_R\overline{\nu}_R^c\nu_R+\text{h.c.} \right)~.
\end{align}
\begin{figure}[t]
    \centering
    \includegraphics[width=0.5\linewidth]{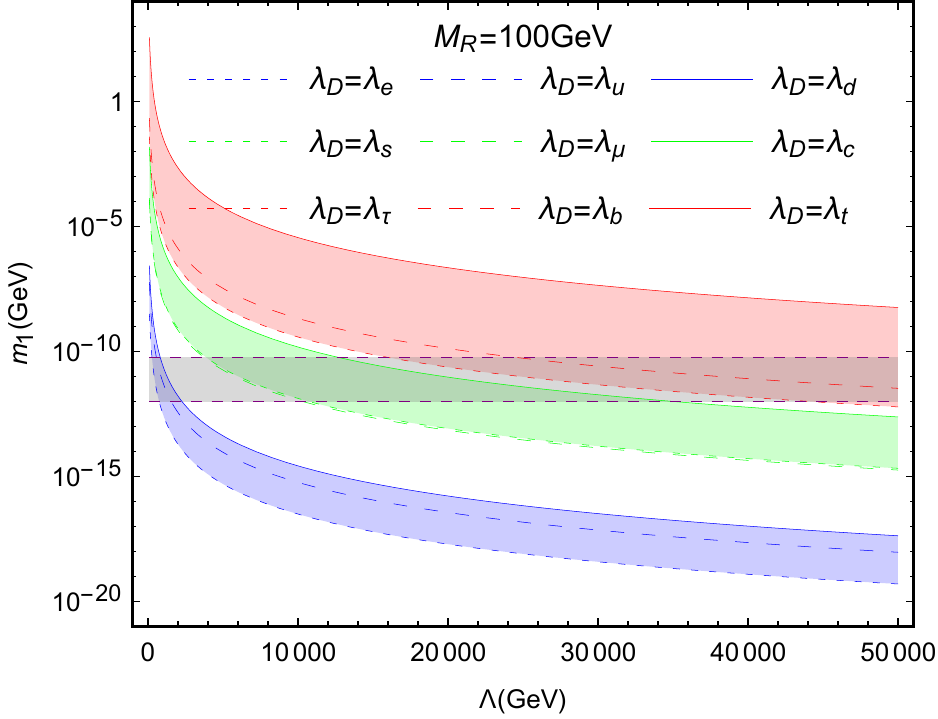}
    \caption{The plot of $m_1$ in Eq.~\eqref{Mrx2} versus the scale $\Lambda$,  where the purple band is the observed neutrino mass range from Eqs.~\eqref{nur1}-\eqref{nur3}.}
    \label{fig:104}
\end{figure}
Applying the seesaw mechanism, the mixing angle is
\begin{align}
    \tan2\theta_\nu=2\frac{\lambda_D v_\text{SM}}{\sqrt{2}M_R}~.
\end{align}
In this case, the small mixing angle condition requires $M_R\gg \lambda_D v_\text{SM}/\sqrt{2}$ so it is suitable to choose the value of $\lambda_D$ in the first and the second generation particles. The active neutrino and heavy neutral lepton masses are given by
\begin{align}\label{Mrx2}
     m_1\simeq \frac{M_h^2 v_{\text{SM}}^2}{8\Lambda^4}\frac{\lambda _{D}^2v_\text{SM}^2}{2  M_R}, ~m_2\simeq \frac{M_h^2 v_{\text{SM}}^2}{8\Lambda^4} M_R~.
\end{align}
Here, $m_2$ can  naturally be smaller than the EW scale by many orders of magnitude, allowing $M_R$ to be in the EW scale 
\begin{align}
    M_R\simeq 100~\text{GeV}~,
\end{align}
and the value of $m_1$ consistently fits with the observed neutrino mass scale in Eqs.~\eqref{nur1}-\eqref{nur3}, shown in figure~\ref{fig:104}. The Yukawa coupling of neutrino can range between  those of the first- and the second-generation particles, with
\begin{align}
300~\text{GeV}\lesssim\Lambda\lesssim40~\text{TeV}~.
\end{align}
Under this constraint, the mass of the mostly right-handed neutrino can be
\begin{align}
   5~\text{eV}\lesssim m_2  \lesssim 2~\text{GeV}~.
\end{align}
Therefore, the mass of the mostly right-handed neutrino in the Type-I seesaw mechanism can extend into the low-energy regime without requiring an unnaturally small $M_R$.
\subsection{Natural large right-handed Majorana mass in the standard Type-I seesaw mechanism}\label{largeMajoranamassKe}
\begin{figure}[t]
    \centering
    \includegraphics[width=0.5\linewidth]{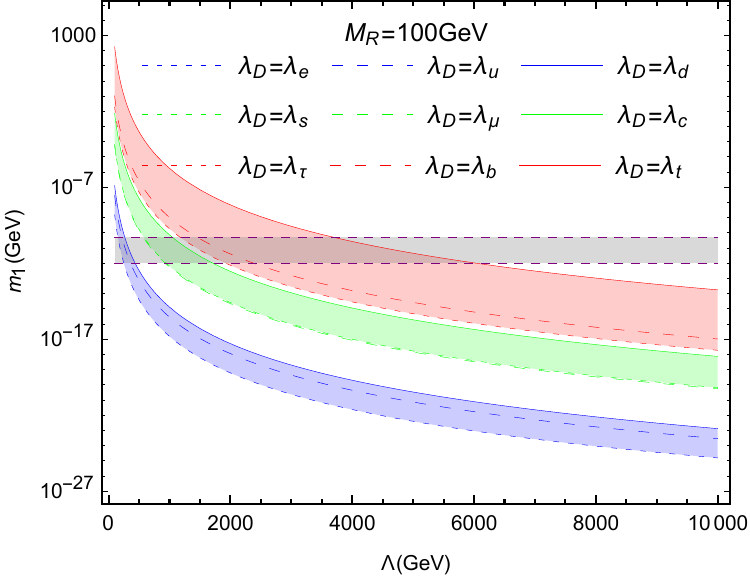}
    \caption{The plot of $m_1$ in Eq.~\eqref{519} versus the scale $\Lambda$,  where the purple band is the observed neutrino mass range from Eqs.~\eqref{nur1}-\eqref{nur3}.}
    \label{fig:active01}
\end{figure}

 We address the problem of unnatural large $M_R$. Without the violation of the gauge invariant, it is feasible to place the kinetic term of the right-handed neutrino both inside and outside the multiplicative Lagrangian, rather than the Majorana mass term as
\begin{align}
        & \mathcal{L}\supset -\overline{L}(\lambda_{D} )\tilde{\phi} ~\nu_{R}+i\overline{\nu}_R\gamma^\mu\partial_\mu \nu_R-M_R\overline{\nu^c_R}\nu_R+\text{h.c.}\nonumber
      \\
      &+\left(\Lambda^4+D_\mu \phi^*D^\mu\phi-i\overline{\nu}_R\gamma^\mu\partial_\mu \nu_R+\overline{L}(\lambda_{D} )\tilde{\phi} ~\nu_{R}+\text{h.c.}\right)e^{-\frac{V}{\Lambda^4}}~,
\end{align} 
where the flavor index is ignored. After the EW phase transition, the right-handed kinetic term gets a non-canonical form multiplied by a constant as
\begin{align}
    \frac{M_h^2 v_\text{SM}^2}{8\Lambda^4}i\overline{\nu}_R\gamma^\mu\partial_\mu \nu_R-\left(\frac{M_h^2 v^2_\text{SM}}{8\Lambda^4}\right)\frac{\lambda_D v_\text{SM}}{\sqrt{2}}\overline{\nu}_L\nu_R-M_R\overline{\nu^c}_R\nu_R+\text{h.c.}~.
\end{align}
 After the field redefinition $\nu_R\to(2\sqrt{2}\Lambda^2 /M_h v_\text{SM}) \nu_R$, the kinetic term is in canonical form. The mass terms of both Dirac and Majorana are scaled up by the parameter $\Lambda$ as
\begin{align}\label{rightmass}
\mathcal{L}_\text{mass}=-\left(\frac{M_h v_\text{SM}}{2\sqrt{2}\Lambda^2}\right)\frac{\lambda_D v_\text{SM}}{\sqrt{2}}\overline{\nu}_L\nu_R-\frac{\Lambda^4}{8M_h^2 v_\text{SM}^2}M_R \overline{\nu^c}_R\nu_R+\text{h.c.}~.
\end{align}
The tree-level pole mass of the right-handed Majorana neutrino, scaled up by the factor $\Lambda^4/M_h^2v^2$, can naturally exceed the EW scale.  
In other words, an unnaturally large scale $M_R$ is not required to raise the active neutrino mass to the eV scale since the kinetic term of $\nu_R$ is incorporated both inside and outside the multiplicative Higgs Lagrangian. By performing the seesaw mechanism,  the mixing angle can be very small
\begin{align}\label{MixingAngle1}
    \text{tan} 2\theta_\nu =  \left(\frac{M_h^3 v_\text{SM}^3}{16\Lambda^6}\right)\frac{\lambda_D v_\text{SM}}{M_R}~,
\end{align}
as it is suppressed by parameter $\Lambda^6$. The mass eigenstates can be written as
\begin{align}\label{519}
     m_1\simeq\left(\frac{M_h^4 v_\text{SM}^4}{128\Lambda^8}\right)\frac{\lambda^2 v_\text{SM}^2}{M_R} , ~m_2\simeq \left(\frac{8 \Lambda^4}{M_h^2 v_{\text{SM}}^2}\right)M_R.
\end{align}
The $m_2$ can naturally exceed the mass scale $M_R$ by many orders of magnitude so it is possible to set $M_R$ around the EW scale, 
\begin{align}
    M_R\sim 100~\text{GeV}~,
\end{align}
and the value of $m_1$ consistently fits with the observed neutrino mass scale in Eqs.~ \eqref{nur1}-\eqref{nur3}, shown in figure.~\ref{fig:active01}. The Yukawa couplings of neutrinos can lie on the Yukawa one of the SM when
the value of $\Lambda$ from the multiplicative Lagrangian is around the EW scale to a few TeV scale
\begin{align}
200~\text{GeV}\lesssim\Lambda\lesssim 7~\text{TeV}~.
\end{align}
 With these results, the mass of the mostly right-handed Majorana neutrino can range as follows,
\begin{align}
    10^3~\text{GeV}\lesssim m_2 \lesssim 10^{9}~\text{GeV}~.
\end{align}
Thus,  the active neutrino mass remains light in the sub-eV regime while the Majorana mass parameter $M_R$ is around the Higgs mass parameter. Consequently, there is no need for an unnaturally large value of the parameter $M_R$ to explain the Type-I seesaw mechanism.  

\subsection{Summary}
The active left-handed neutrino mass $m_1$ can naturally be obtained in the sub-eV scale with $\lambda_D\sim\lambda^\text{SM}_f$ and $M_R\simeq 100$ GeV. 
The mostly right-handed neutrino mass $m_2$ can be spanned several orders of magnitude depending on the Majorana mass term and the kinetic energy
\begin{align}
        & \mathcal{L}\supset ai\overline{\nu}_R\gamma^\mu\partial_\mu \nu_R-c\overline{L}(\lambda_{D} )\tilde{\phi} ~\nu_{R}-bM_R\overline{\nu^c_R}\nu_R+\text{h.c.}\nonumber
      \\
      &+\left(\Lambda^4+D_\mu \phi^*D^\mu\phi-Ai\overline{\nu}_R\gamma^\mu\partial_\mu \nu_R+C\overline{L}(\lambda_{D} )\tilde{\phi} ~\nu_{R}+BM_R\overline{\nu^c_R}\nu_R+\text{h.c.}\right)e^{-\frac{V}{\Lambda^4}},
\end{align} 
where $a,~b,~c,~A,~B,~C=0$ or $1$. This framework naturally yields three distinct regimes of $m_2$, as follows.
\begin{itemize}
\item Standard Type-I seesaw ($A=B=C=0$ and $a=b=c=1)$: 
    \\
    $M_R\simeq 100$ GeV, $m_2\simeq 100$ GeV, $\lambda_D\lesssim \lambda_{f,\text{1st}}$.

    \item  Natural EW $m_2$ ($A=B=0$ and $a=b=c=C=1)$: 
    \\
    $M_R\simeq 100$ GeV, $m_2\simeq 100$ GeV, $\lambda_{f,\text{1st}}\lesssim\lambda_D\lesssim \lambda_{f,\text{3rd}}$ with $200~\text{GeV}\lesssim\Lambda\lesssim 7$~TeV.

    \item Natural small $m_2$ ($A=0$ and $a=b=B=c=C=1$): 
    \\
    $M_R\simeq 100$~GeV, $5~\text{eV}\lesssim m_2 \lesssim 2~\text{GeV}$, $\lambda_{f,\text{1st}}\lesssim\lambda_D\lesssim \lambda_{f,\text{2nd}}$ with $300~\text{GeV}\lesssim\Lambda\lesssim 40$~TeV.

    \item Natural large $m_2$ ($B=0$ and $a=A=b=c=C=1)$: 
    \\
     $M_R\simeq 100$~GeV, $10^4~\text{GeV}\lesssim m_2 \lesssim 10^{12}~\text{GeV}$,  $\lambda_{f,\text{1st}}\lesssim\lambda_D\lesssim \lambda_{f,\text{3rd}}$ with $200~\text{GeV}\lesssim\Lambda\lesssim7$ TeV.
\end{itemize}
In this model, the dimensionless ratios constructed from the dimensionful parameters are
\begin{align}
    10^{-3}\lesssim\left(\frac{M_R}{\Lambda},~\frac{M_R}{v},~\frac{\mu}{M_R},~\frac{\mu}{v},~\frac{v}{\Lambda},~\frac{\mu}{\Lambda}\right)\lesssim 1,
\end{align}
whereas, in the traditional SM+Type-I seesaw mechanism, the dimensionless ratio $M_R/v$ can range within $1$ to $10^{15}$. This framework thus reduces the widely separated scale of $M_R$ and $v$ by approximately ten orders of magnitude.

\section{Discussion}\label{section5}

\subsection{Dirac neutrino constraint}
Firstly, we discuss the constraint on the Dirac neutrino model. We check the consistency of our Dirac neutrino model with the four neutrino effective interactions
 \begin{align}\label{4neutrinointeraction}
   &\mathcal{L}_{eff}= G_S \overline{\nu}_L\nu_R\overline{\nu}_L\nu_R+G_S \overline{\nu}_R\nu_L\overline{\nu}_R\nu_L+\tilde{G}_S \overline{\nu}_L\nu_R\overline{\nu}_R\nu_L~.
 \end{align}
These effective coupling constants are directly constrained by the current measurement of the effective number of relativistic neutrino species $N_{eff}$\cite{Luo:2020sho,Masso:1994ww}
\begin{align}\label{lowerlimitofGS}
     \sqrt{1/G_S}>12.4~\text{TeV},~ \sqrt{1/\tilde{G}_S}>8.1~\text{TeV}.
 \end{align}
 In our proposed model, these effective interactions can be obtained by integrating out of the Higgs particle in the intermediate state from the scattering process $\nu\nu\to\nu\nu$. We have
 \begin{align}
    \mathcal{L}_{eff}= \frac{m_{\nu_i}^2}{M_h^2v_\text{SM}^2} (\bar{\nu}_L\nu_R+\bar{\nu}_R\nu_L)^2~,
 \end{align}
 which gives $G_S$ and $\tilde{G}_S$ in the same order of magnitude as 
 \begin{align}\label{GSprediction}
     G_S=\frac{m_{\nu_i}^2}{M_h^2v_\text{SM}^2}, \tilde{G}_S=2\frac{m_{\nu_i}^2}{M_h^2v_\text{SM}^2}~.
 \end{align}
 Substituting Eqs.~\eqref{nur1}-\eqref{nur3} into \eqref{GSprediction}, the parameter $1/\sqrt{G^i_S}$ for the neutrino in the $i^\text{th}$ generation can be possible in the regime
 \begin{align}
     10^{11} ~\text{TeV}\lesssim1/\sqrt{G^i_S}\lesssim 10^{13}~\text{TeV}~,
 \end{align}
 which is consistent with the lower bounded limit in Eq.~\eqref{lowerlimitofGS}.

 \subsection{Heavy Dirac mass and third generation particle}\label{third}
\begin{figure}[t]
    \centering
    \subfigure[]{\includegraphics[width=0.48\linewidth]{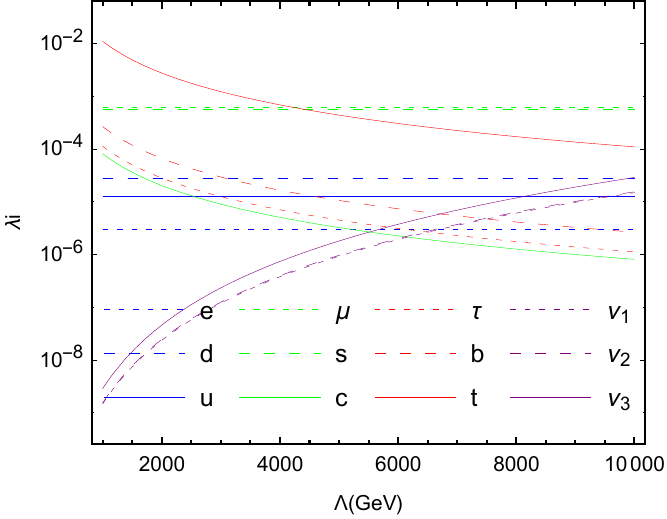}}
    \subfigure[]{\includegraphics[width=0.48\linewidth]{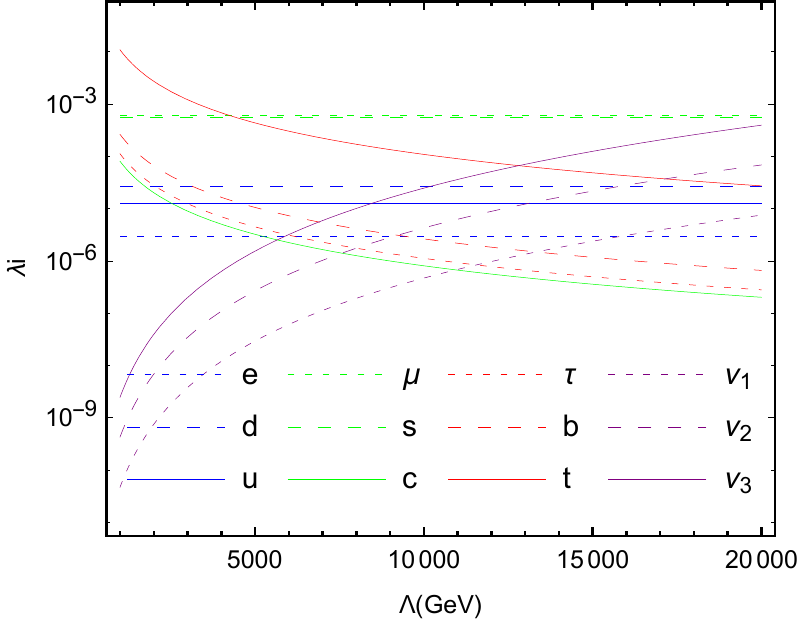}}
    \caption{Plot of Yukawa coupling in Eqs.~\eqref{98}-\eqref{100} versus the scale $\Lambda$ in GeV unit with (a) $\sum m_{\nu_i}=0.12$ eV and (b) $\sum m_{\nu_i}=0.06$ eV. }
    \label{fig:fy}
\end{figure}

\begin{figure}[t]
     \centering
     \subfigure[]{\includegraphics[width=0.45\linewidth]{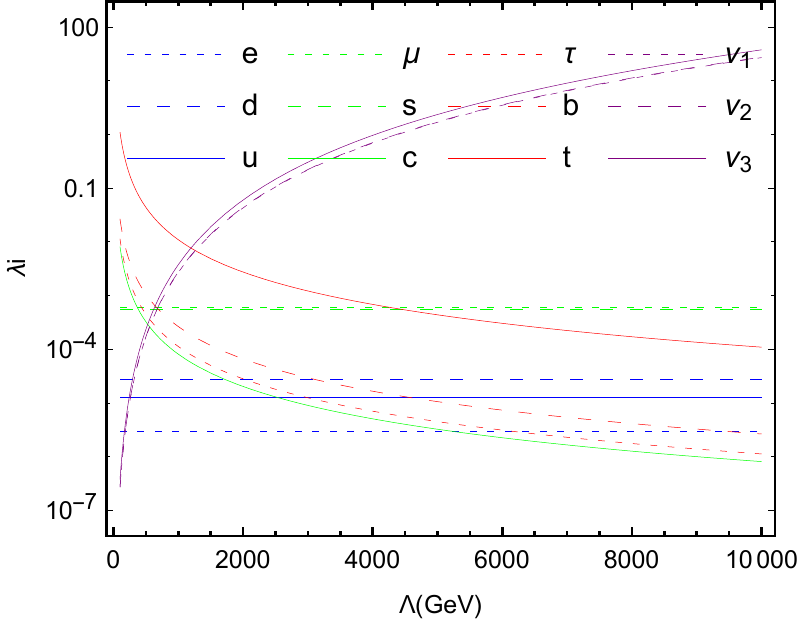}}
     \subfigure[]{\includegraphics[width=0.45\linewidth]{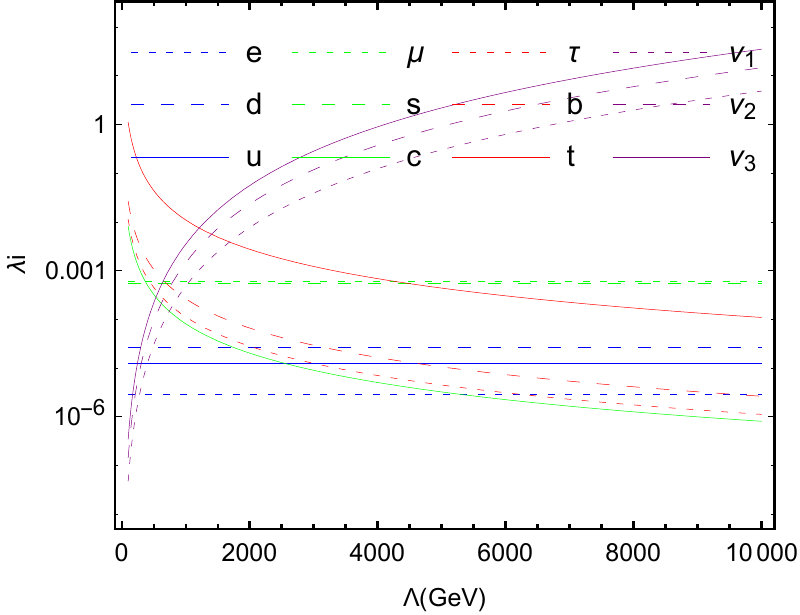}}
     \subfigure[]{\includegraphics[width=0.45\linewidth]{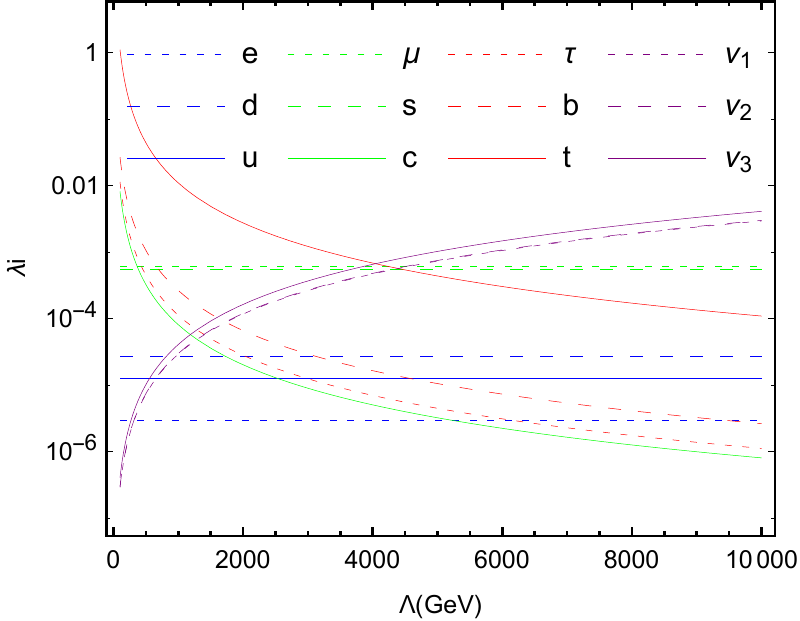}}
     \subfigure[]{\includegraphics[width=0.45\linewidth]{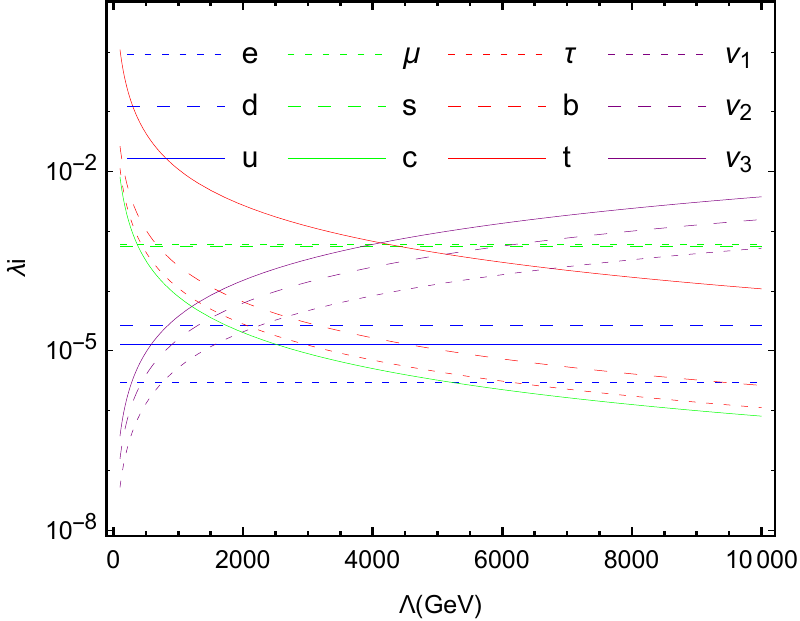}}
     \subfigure[]{\includegraphics[width=0.45\linewidth]{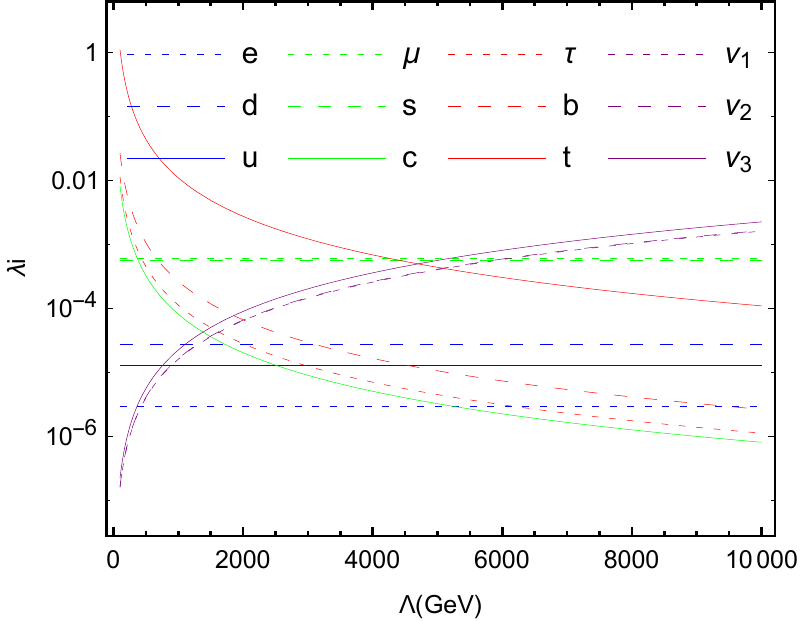}}
     \subfigure[]{\includegraphics[width=0.45\linewidth]{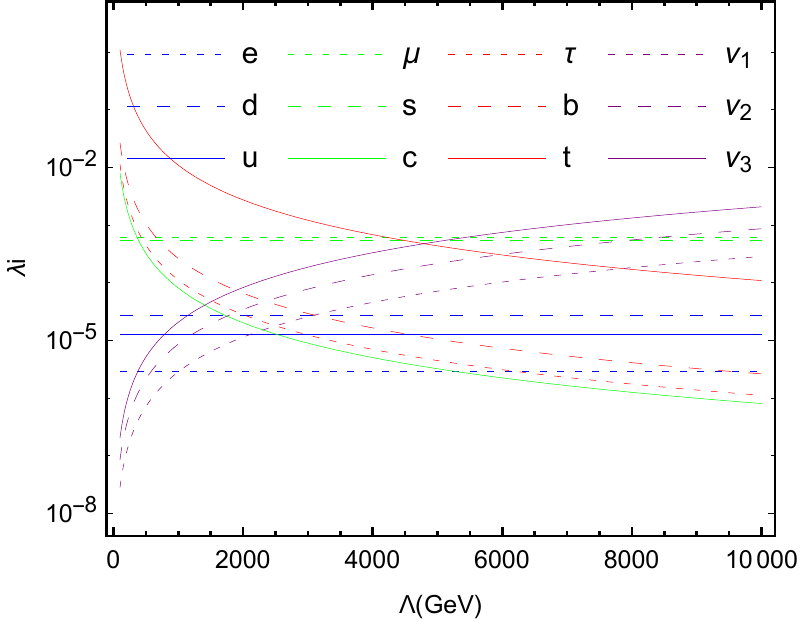}}
     \caption{(a) and (b) are the plot of Yukawa couplings from Eqs.~\eqref{99}-\eqref{100},  Eq.~\eqref{619} with $M_R=100$~GeV, where (a) $\sum m_\nu=0.12$~eV, and (b) $\sum m_\nu=0.06$~eV. (c) and (d) are the plot of Yukawa couplings from Eqs.~\eqref{99}-\eqref{100},  Eq.~\eqref{620} with $M_R=100$~GeV,  where (c) $\sum m_\nu=0.12$~eV, and (d) $\sum m_\nu=0.06$~eV.
     (e) and (f) are the plot of Yukawa couplings from Eqs.~\eqref{99}-\eqref{100},  Eq.~\eqref{620} with $M_R=30$~GeV,  where (e) $\sum m_\nu=0.12$~eV, and (f) $\sum m_\nu=0.06$~eV.}
     \label{fig:8888}
 \end{figure}

  In this discussion, we study the mechanism described by Eq.~\eqref{mechanism} that influences the Dirac mass, where $\hat{A}$ is the kinetic term of Dirac fermion. 
  This exploration is motivated in section.~\ref{largeMajoranamassKe}, which addresses the issue of an unnaturally large Majorana mass by incorporating the right-handed part of the kinetic term inside and outside the multiplicative Lagrangian.  To simplify our analysis, since our primary focus is on the fermion mass scale, we bypass the complexity of the full SM gauge group by considering the toy model of a multiplicative complex scalar field coupled with the fermion as
  \begin{align}
&\mathcal{L}=i\overline{\Psi}_L\gamma^\mu\partial_\mu\Psi_L+i\overline{\Psi}_R\gamma^\mu\partial_\mu\Psi_R\nonumber
      -\frac{\lambda_\Psi}{\sqrt{2}} (\phi\overline{\Psi}_L\Psi_R+h.c.)\nonumber
      \\
      &+(-\Lambda^4+\partial_\mu \phi^\dagger\partial^\mu\phi+ i\overline{\Psi}_R\gamma^\mu\partial_\mu\Psi_R
      )e^{\frac{V}{\Lambda^4}}~.
  \end{align}
  We will draw an analogy between $\phi$ and $\Psi$ as the Higgs and fermion in the SM, respectively, where $v=246$GeV, $\mu=88$GeV. Furthermore, because the choice of the parameter $\epsilon$ does not affect the energy scale of the fermion mass, as shown in the previous section,  we will consider only the choice $\epsilon=-1$. After the spontaneous symmetry breaking (SSB), the mass term of $\Psi$ is
   \begin{align}\label{masstermPsi}
       \frac{\lambda_\Psi v}{\sqrt{2}}\overline{\Psi}_R\Psi_L+\text{h.c.}~,
   \end{align}
and the kinetic energy of fermion is in non-canonical form
   \begin{align}\label{kePsi1}
       \left(\frac{M_h^2 v^2}{8\Lambda^4}\right)i\overline{\Psi}_R\gamma^\mu\partial_\mu\Psi_R+i\overline{\Psi}_L\gamma^\mu\partial_\mu\Psi_L~.
   \end{align}
To reorganize the kinetic term into a renormalized canonical one, we can perform the field redefinition 
   \begin{align}\label{redPsi}
     \Psi_R\to \left(\frac{8\Lambda^4}{M_h^2 v^2}\right)^{\frac{1}{2}}\Psi_R~.
   \end{align}
The mass term in Eq.~\eqref{masstermPsi} then is reorganized into the form
   \begin{align}\label{610}
      \left(\frac{8\Lambda^4}{M_h^2 v^2}\right)^{\frac{1}{2}}\frac{\lambda_\Psi v}{\sqrt{2}}\overline{\Psi}_R\Psi_L+\text{h.c.}~.
   \end{align}
From Eq.~\eqref{610}, the expression of the fermion mass is
   \begin{align}
       m_{\Psi}=\left(\frac{8\Lambda^4}{M_h^2 v^2}\right)^{\frac{1}{2}}\frac{\lambda_\Psi v}{\sqrt{2}}~.
   \end{align}
   The result is different compared to the case of the Majorana mass that is scaled by  $\Lambda^4$ shown in Eq.~\eqref{rightmass}. From the constraint of $\Lambda$ in section.~\ref{section4}, $5.7~\text{TeV}<\Lambda<88~\text{TeV}$, $m_\Psi$ can be larger than the normal fermion mass, $m_f=\lambda_f v/\sqrt{2}$, from the standard Higgs mechanism with
   \begin{align}\label{ratio31}
       \frac{m_\Psi}{m_f}\sim 10^3-10^6~.
   \end{align}
This scale is coincident with the mass scale of the third- and first-generation particles,
   \begin{align}\label{ratio312}
       \frac{m_{f,\text{3rd}}}{m_{f,\text{1st}}}\sim 10^{3}-10^{5}~,
   \end{align}
   for example $m_t/m_e\simeq 3.46\times 10^5$, $m_t/m_u\simeq 7.86\times 10^4$, and $m_\tau/m_e\simeq 3.44 \times 10^3$. From Eq.~\eqref{ratio31} and Eq.~\eqref{ratio312}, we may hypothesize that the mass hierarchy of the third-generation fermions can arise from the differences in the configuration of kinetic and Yukawa terms in the Lagrangian, leading to differences in the mass scale
   \begin{align}
       &(m_{\nu_1},m_{\nu_2},m_{\nu_3})\sim \left(\frac{M_h^2 v^2}{8\Lambda^4} \right)\frac{\lambda v}{\sqrt{2}}~, \label{98}
       \\
       &(m_e,m_u,m_d,m_s,m_\mu)\sim \frac{\lambda v}{\sqrt{2}}\label{99}~,
       \\
       &(m_{\tau}, m_c, m_b, m_t)\sim \left(\frac{8\Lambda^4}{M_h^2 v^2}\right)^{\frac{1}{2}} \frac{\lambda v}{\sqrt{2}}\label{100}~,
   \end{align}
   where we have assumed that neutrino is Dirac type.
With this hypothesis, we find that the size of the Yukawa coupling of all SM fermions can be compactified within the range $10^{-4}-10^{-6}$($10^{-3}-10^{-6}$), at $\Lambda\simeq 6$~TeV($\Lambda\simeq 11$~TeV) for $\sum m_{\nu_i}=0.12$ eV (for $\sum m_{\nu_i}=0.06$ eV), shown in figure~\ref{fig:fy}.
 Hence, the dimensionless ratios between the Yukawa couplings
naturally lie within the $\mathcal{O}(1)$ of the Dirac naturalness, 
\begin{align}\label{naturalness}
    10^{-2}(10^{-3})\lesssim\lambda_{f_i}/\lambda_{f_j}\lesssim 1~,
\end{align}
 where $\lambda_{f_i}<\lambda_{f_j}$.  The fermion mass, spanning between $10^{-14}$GeV to $100$GeV, can be explained by the model that has the dimensionless ratios of parameters within the range
 \begin{align}
     10^{-2}\lesssim \left(\frac{\mu}{\Lambda},\frac{v}{\Lambda},\frac{\mu}{v}\right)\lesssim 1~,
 \end{align}
 without extending any internal symmetry group, or introducing a new scalar boson. On the other hand, if neutrinos are Majorana type in the case $m_2>100$ GeV, the Yukawa couplings of the mostly left-handed neutrinos in both cases (case $A=B=0$, $a=b=c=C=1$ and case $B=0$, $a=A=b=c=C=1$) take the same form 
 \begin{align}\label{619}
     \lambda_{\nu_i}=\frac{8  \Lambda ^4 \sqrt{2M_R m_{\nu _i}} }{M_h^2 v_{\text{SM}}^3}~.
 \end{align}
 To obtain $m_{\nu_i}$ within the range of Eqs.  \eqref{nur1}-\eqref{nur3}, the Yukawa couplings of neutrinos are larger than those of the charged fermions by many orders of magnitude, shown in figure.~\ref{fig:8888}(a) and figure.~\ref{fig:8888}(b). Therefore, the case of a large Majorana mass, $m_2>100$ GeV, cannot be reconciled with the explanation for the third generation.
 Only the natural small $m_2<100$~GeV (case $A=0$, $a=b=B=c=C=1$) can be incorporated together with the explanation of the third generation. In this case, the Yukawa couplings of neutrinos are given by 
 \begin{align}\label{620}
     \lambda_{\nu_i}=\frac{4 \Lambda ^2 \sqrt{M_Rm_{\nu _i}} }{M_h v_{\text{SM}}^2}~.
 \end{align}
The size of Yukawa interaction both the mostly left-handed Majorana neutrinos and the charged fermions can be compacted in the scale $10^{-4}-10^{-6}$ at $\Lambda\simeq 4$~TeV, shown in figure~\ref{fig:8888}(c) and figure~\ref{fig:8888}(d). In this case, we obtain $m_2\simeq 41$~keV, which is not low enough to incorporate in the warm dark matter problem. To reduce $m_2$ in to a few kev scale, one just needs $M_R\simeq 30$~GeV, and $\Lambda\simeq 4.7$~TeV to get $m_2\simeq7$~keV while all of Yukawa couplings are around the order $10^{-4}-10^{-6}$, shown in figure~\ref{fig:8888}(e) and figure~\ref{fig:8888}(f). Here, all of the dimensionless ratios of parameters in the model can fall within the range
\begin{align}
    10^{-2}\lesssim \left(\frac{M_R}{\Lambda},\frac{M_R}{v},\frac{M_R}{\mu},\frac{\mu}{\Lambda},\frac{v}{\Lambda},\frac{\mu}{v},\frac{\lambda_{f_i}}{\lambda_{f_j}}\right)\lesssim 1~,
\end{align}
 to explain the mass spectrum of particles spanning 
between meV to 100 GeV scale.

 \subsection{Validity of model}
 We discuss the UV-cutoff energy ($\Lambda_\text{UV}$) of the non-renormalizable multiplicative Higgs Lagrangian, which can be specified from the inverse coefficient of the non-renormalized operators as
 \begin{align}
     \frac{\hat{O}_n}{(\Lambda_{(n)})^{n-4}}~,
 \end{align}
 where $\hat{O}_n$ is an operator with mass dimension $n>4$, and $\Lambda_{(n)}$ is the cutoff parameter. In brief, this type of operator can generally yield a scattering amplitude in the form $\mathcal{A}_n\sim E/\Lambda_{(n)}$, where $E$ is the energy of the particle. When $E>\Lambda_{(n)}$, the perturbativity and unitarity of $\mathcal{A}_n$ break down ($\mathcal{A}_n>1$). As an effective field theory, the model is valid up to the energy scale $E_\text{maximum}\sim \Lambda_{(n)}$. Hence, the UV-cutoff energy of particle in the model should be $\Lambda_\text{UV}\sim \Lambda_{(n)}$. If the model includes multiple cutoff parameters, as in \cite{Bezrukov:2010jz},  $\Lambda_\text{UV}$ is determined by the minimum values in the set of $\Lambda_{(n)}$, given by $\Lambda_\text{UV}=\text{min}[\Lambda_{(n)}]$.
 
 In the multiplicative Higgs model,  a series of non-renormalized operators $\chi^{n}$ with $n>4$ can be obtained from the series expansion of the exponential potential $V_\phi$ in Eq.~\eqref{Veff},
 \begin{align}
     V_\phi=&-\epsilon\Lambda^4e^{-\frac{-\mu^2\phi^\dagger\phi+\lambda (\phi^\dagger\phi)^2}{\epsilon\Lambda^4}}+\epsilon\Lambda^4\nonumber~,
     \\
     =&-\epsilon\Lambda^4 e^{ -\frac{\chi ^4 M_h^2}{8 \Lambda ^4 v^2 \epsilon }+\frac{v^2 M_h^2}{8 \Lambda
   ^4 \epsilon }-\frac{\chi ^3 M_h^2}{2 \Lambda ^4 v \epsilon }-\frac{\chi
   ^2 M_h^2}{2 \Lambda ^4 \epsilon }}+\epsilon\Lambda^4~,\nonumber
   \\
   =&-\epsilon\Lambda^4 e^{\frac{v^2 M_h^2}{8 \Lambda
   ^4 \epsilon }-\frac{\chi^2}{\epsilon\Lambda_{2\chi}^2}-\frac{\chi^3}{\epsilon\Lambda_{3\chi}^3}-\frac{\chi^4}{\epsilon\Lambda_{4\chi}^4}}+\epsilon\Lambda^4~. \label{vhiggs}
\end{align}
 From Eq.~\eqref{vhiggs}, the terms $\chi^2$, $\chi^3$, and $\chi^4$ on the exponential function are divided by the different energy scales,
 \begin{align}
  &\Lambda_{2\chi}=\left(\frac{\sqrt{2}\Lambda}{M_h}\right)\Lambda,~\Lambda_{3\chi}=\left(\frac{2v\Lambda}{M_h^2}\right)^\frac{1}{3}\Lambda,~\Lambda_{4\chi}=\left(\frac{\sqrt{8}v}{M_h}\right)^\frac{1}{2}\Lambda~.
 \end{align}
 If $\Lambda>M_h$, and $\Lambda>v$, it is easy to see that $\Lambda_{2\chi}>\Lambda_{3\chi}>\Lambda_{4\chi}$ since the dimensionless coefficients in the bracket of $\Lambda_{2\chi}$, $\Lambda_{3\chi}$, $\Lambda_{4\chi}$  are proportional to $\Lambda^1$, $\Lambda^{1/3}$, $\Lambda^0$, respectively. Therefore, the value of $\Lambda_\text{UV}=\text{min}[\Lambda_{n}]$ can be determined by the non-renormalized operators arising  from the series expansion of $\Lambda^4\exp(\chi^4/\Lambda_{4\chi}^4)$, as
 \begin{align}\label{UVchi}
     \Lambda_\text{UV}\simeq \left(\frac{\sqrt{8}v}{M_h}\right)^\frac{1}{2}\Lambda~\sim ~2.4\Lambda~.
 \end{align}
 If we consider the UV-cutoff in terms of $h$ field, the value of $\Lambda_\text{UV}$ differs from Eq.~\eqref{UVchi} by the factor $\exp(-M_h^2v^2/8\Lambda^4)$ coming from the field redefinition in Eq.~\eqref{redchi}. This factor is very close to 1 when $\Lambda\gg 100$~GeV. Thus, the UV cutoff between the $\chi$ and $h$ fields is approximately equivalence. For the theory of neutrino with $\Lambda\sim 4~\text{TeV}-100$~TeV, $\Lambda_\text{UV}$ is around $10~\text{TeV}-200$~TeV.
 With the $4\pi$ counting \cite{Manohar:1983md,Gavela:2016bzc}, which is the highest momentum circulating in the loop, $ \Lambda_\text{UV}$ can reach around $4\pi(2.4)\Lambda\simeq 100~\text{TeV}-1000$ TeV. 

\subsection{Open problems}
It is important to emphasize that although this framework can reduce the widely separated scale of the Yukawa couplings to a hundred scale, the flavor problems in contexts of the origin of particle mixing and the hierarchy of fermion mass still remain.  The  $\lambda_{f_i}$ are still the unpredictable variables, which require an alignment with experimental observations.  Moreover, the PMNS matrix elements remain unpredictable values within our current framework.  The incomprehensible result of an empirical Koide formula \cite{KOIDE1983161,SUMINO2009477,Rodejohann:2011jj} is still enigmatic without theoretical support. This framework may be a first step toward exploring the flavor problem and could provide insights for further study.

For example,  from the figure~\ref{fig:fy} in section~\ref{third}, the overall hierarchies come from the contribution of $t$, $\mu$, and $s$, see the solid red line and the green dashing line, since the scale factor in Eq.~\eqref{100} is around $10^3-10^4$ for $\Lambda\simeq 6~\text{TeV}-11$~TeV. It is insufficient to reduce the scale of the top quark down to the first-generation particle. By excluding $\lambda_{\mu,s,t}$ in figure~\ref{fig:fy}, the size of dimensionless ratio can be compacted within the range $10^{-1}\lesssim\lambda_{f_i}/\lambda_{f_j}\lesssim 1$. We may hypothesize that at least two mass generations proportional to $\Lambda^1$ and $\Lambda^{3}$ might be embedded within the multiplicative Lagrangian to generate the muon, strange quark, and top quark mass scale 
\begin{align}\label{114}
  m_{\mu,s}\sim\left(\frac{8\Lambda^4}{M_h^2v^2}\right)^\frac{1}{4}m_{f,\text{1st}},~ m_{t}\sim\left(\frac{8\Lambda^4}{M_h^2v^2}\right)^\frac{3}{4}m_{f,\text{1st}}~,
\end{align}
which are not found in this paper. Under this hypothesis, all ratios of the Yukawa couplings can then close the gap to the natural scale between  
\begin{align}
    \frac{\lambda_{f_i}}{\lambda_{f_j}}\simeq\frac{g_i}{g_j}\sim 0.1-1~,
\end{align}
which is consistent with the ratios between the gauge couplings in the SM.
Alternatively, it is plausible to hypothesize that the mass scale of $\mu,s,t$ may originate from a different mechanism rather than generated from the multiplicative Lagrangian. For example, the non-diagonal mass matrix of the charged lepton is allowed with the current framework of the neutrino oscillation since the PMNS matrix $U_\text{PMNS}=U_l^\dagger U_\nu$ can be obtained by rotating the charged lepton and neutrino field into the mass eigenstate with the unitary matrices $U_l$ and $U_\nu$. We exemplify a minimal modification of the charged lepton mass matrix as
\begin{align}\label{massmatrixl}
    M_l=\frac{v}{\sqrt{2}}\begin{pmatrix}
        \lambda_{e} & 0 & 0
        \\
        0 & \frac{2\sqrt{2}\Lambda^2}{M_h^2v^2} \lambda_\tau & \frac{2\sqrt{2}\Lambda^2}{M_h^2v^2} \lambda_{\mu\tau}
        \\
        0 & \frac{2\sqrt{2}\Lambda^2}{M_h^2v^2} \lambda_{\mu\tau} & \frac{2\sqrt{2}\Lambda^2}{M_h^2v^2} \lambda_{\tau}
    \end{pmatrix}~,
\end{align}
in which the number of parameters in our theory is still the same. 
At $\Lambda=6$ TeV, the Yukawa couplings can be naturally the same order of magnitude $\lambda_e=2.94\times 10^{-6}$, $\lambda_{\mu\tau}=1.45\times 10^{-6}$, $\lambda_\tau=1.63\times 10^{-6}$ which can provide three mass eigenstates corresponding to the mass of charged leptons $(0.511,~105,~1776)$MeV. In this point of view, the kinetic terms of muon and tau are written in the same scheme leading to the mass terms of both particles proportional to $\Lambda^2$ so the new mass scales from the multiplicative Lagrangian in Eq.~\eqref{114} are not essential to provide the Yukawa coupling falling in the same order of magnitude. 

 Finally, although the hidden structure in Eq.~\eqref{114} can be found, the fine-tuning problem still persists. The absolute values of $\lambda_f$  remain around $10^{-6}$, which is still 5 orders of magnitude smaller than those of the gauge couplings. A comprehensive reconfiguration of the model may thus be necessary, for example,
 \begin{align}
       &(m_e,m_u,m_d)\sim \left(\frac{M_h^2 v^2}{8\Lambda^4} \right)\frac{\lambda v}{\sqrt{2}}~, \label{999}
       \\
       &(m_{\tau}, m_c, m_b,m_s,m_\mu)\sim \left(\frac{M_h^2 v^2}{8\Lambda^4} \right)^{\frac{1}{2}} \frac{\lambda v}{\sqrt{2}}~,\label{9999}
       \\
       &m_t\sim \frac{\lambda v}{\sqrt{2}}\label{99999}~.
   \end{align}
For $\Lambda=1500$ GeV, the Yukawa couplings of the charged fermion are at a natural scale, 
\begin{align}
    &\lambda_e=0.126,~\lambda_d=1.16,~~\lambda_u=0.542, \nonumber
    \\
    &\lambda_\mu=0.125,~\lambda_s=0.115,~\lambda_c=1.46,\nonumber
    \\
    &\lambda_\tau=2.11,~~~\lambda_b=4.97,~~\lambda_t=0.995,\nonumber
\end{align}
as the gauge couplings. Here, these values are actually within $\mathcal{O}(1)\sim 0.1- 1$ rather than the approximate range of $\mathcal{O}(1)\sim 0.001- 1$ used in our analysis.  However, the price to pay is that the origin of the Dirac neutrino mass must be abandoned. Consequently, the seesaw mechanism is essential to generate the neutrino mass scale.

\section{Conclusion}\label{lastsection}

The Lagrangian is modeled in a minimalistic way without the addition of a new scalar field and extending the gauge group. Only one dimensionful parameter, with a value close to the electroweak scale, is introduced into the SM model to explain the widely separated scale between the neutrinos and the charged fermions. 

The model is based on two main hypotheses: 
\begin{enumerate}
   \item The Higgs Lagrangian takes a nonstandard form. 
   \item The Yukawa interaction and fermion kinetic energy can be included in the multiplicative Lagrangian in a variety of gauge-invariant configurations. 
\end{enumerate}
 With these hypotheses, the hierarchy structure of the Dirac fermion can be generated by the dimensionless factor 
\begin{align}\label{fc}
   m_{f}(N)= \left(\frac{M_h^2v_\text{SM}^2}{8\Lambda^4}\right)^N\frac{\lambda_f v_\text{SM}}{\sqrt{2}}~,
\end{align}
where $N=+1,+1/2,~0,-1/2$ depending on the schemes, shown in the section~\ref{DiracNeutrino}, section~\ref{SmallMajoranamassKe}, section~\ref{largeMajoranamassKe}, and section~\ref{third}. Additionally, the hierarchy of the Majorana mass is also generated across various scales 
\begin{align}
    M_R(n)=\left(\frac{M_h^2v_\text{SM}^2}{8\Lambda^4}\right)^n M_R~,
\end{align} 
where $M_R$ is hypothesized around the electroweak scale, and $n=-1,0,+1$, depending on the schemes, shown in  section~\ref{SmallMajoranamassKe} and section~\ref{largeMajoranamassKe}. With these hierarchies, the neutrino mass scale can naturally be produced at the meV scale when \(N = +1\), which is applicable to both the Dirac and Majorana neutrino types. Moreover, these mass scales have the potential to address the flavor problem and shed light on the keV dark matter. The third-generation mass scale can be derived from \(m_f(1/2)\) with \(\Lambda = 4~\text{TeV}-6\)~TeV, and a mostly right-handed Majorana neutrino at the keV scale can be obtained naturally from \(M_R(+1)\) with \( M_R \approx 100/3 \) GeV.
\appendix
\section{Neutrino mass range (IH)}\label{IHsec}
The constraints on the neutrino mass difference in the inverse hierarchy are
\begin{align}\label{constraint2}
 \sum m_{\nu_i}& <0.12~\text{eV},\nonumber
 \\
     \Delta m_{21}^2&\simeq 7.53\times 10^{-5}~\text{eV}^2~,\nonumber
    \\
    \Delta m_{32}^2&\simeq- 2.529\times 10^{-3}~\text{eV}^2~.
\end{align}
We find that, when $\sum m_{\nu_i}<0.1$ eV, it leads to the negative  neutrino mass $m_{\nu_i}$. Under the condition $ \sum m_{\nu_i}=0.1~\text{eV}-0.12$ eV,
the neutrino masses $m_{\nu_i}$ can be  given by
\begin{align}
  0.050~\text{eV}\lesssim  &m_{\nu_1}\lesssim 0.052~\text{eV}~,\label{nur1I}
  \\
  0.050~\text{eV}\lesssim  &m_{\nu_2}\lesssim 0.053~\text{eV}~,\label{nur2I}
  \\
  0.00018~\text{eV}\lesssim  &m_{\nu_3}\lesssim 0.016~\text{eV}~.\label{nur3I}
\end{align}
This then provides the Yukawa coupling in the regime
\begin{align}
    8.5\times 10^{-24}\frac{\Lambda^4}{\text{GeV}^4}<\lambda_{\nu_i}< 2.6\times 10^{-21}\frac{\Lambda^4}{\text{GeV}^4}~,
\end{align}
which is approximately around the NH case, shown in Eq.~\eqref{nuYrange1}.

\acknowledgments

This research has received funding
support from the NSRF via the Program Management Unit
for Human Resources \& Institutional Development, Research
and Innovation [grant number B37G660013].

\bibliographystyle{JHEP}
\bibliography{mybib.bib}
\end{document}